\documentclass[useAMS,usenatbib,usegraphicx]{mn2e}

\def\phn{\phantom{0}}       
\def\phd{\phantom{.}}       

\def\lesssim{\mathrel{\hbox{\rlap{\hbox{\lower4pt\hbox{$\sim$}}}\hbox{$<$}}}}
\def\gtrsim{\mathrel{\hbox{\rlap{\hbox{\lower4pt\hbox{$\sim$}}}\hbox{$>$}}}}

\newcommand{\mamo}[1]{\mbox{$#1$}}
\newcommand{\unit}[1]{\ifmmode \:\mbox{\rm #1}\else \mbox{#1}\fi}
\newcommand{\sbr}[1]{_{\rm #1}}
\newcommand{\spr}[1]{^{\rm #1}}
\newcommand{\msun}{\mamo{M_{\odot}}}
\newcommand{\secref}[1]{Section~\ref{sec:#1}}

\newcommand{\eqref}[1]{equation~(\ref{eq:#1})}

\newcommand{\figref}[1]{Fig.~\ref{fig:#1}}
\newcommand{\tabref}[1]{Table~\ref{tab:#1}}

\title[]{CFHTLenS: The Environmental Dependence of Galaxy Halo Masses from Weak Lensing}
\author[Bryan~R.~Gillis {\it et~al.}]{Bryan~R.~Gillis$^{1}$\thanks{E-mail:
bgillis@uwaterloo.ca}, Michael~J.~Hudson$^{1,2}$,
Thomas~Erben$^3$,
Catherine~Heymans$^4$, \newauthor
Hendrik~Hildebrandt$^{3,5}$,
Henk~Hoekstra$^{6,7}$,
Thomas~D.~Kitching$^4$,
Yannick~Mellier$^8$,  \newauthor
Lance~Miller$^9$,
Ludovic~van Waerbeke$^5$,
Christopher~Bonnett$^{10}$,
Jean~Coupon$^{11}$, \newauthor
Liping~Fu$^{12}$,
Stefan~Hilbert$^{13,3}$,
Barnaby~T.P.~Rowe$^{15,26}$, \newauthor
Tim~Schrabback$^{3,13,6}$,
Elisabetta~Semboloni$^{6}$,
Edo~van Uitert$^{6,3}$, \newauthor
and 
Malin~Velander$^{9,6}$.
\\
$^{1}$Department of Physics and Astronomy, University of Waterloo, Waterloo, ON N2L 3G1, Canada.\\
$^{2}$Perimeter Institute for Theoretical Physics, 31 Caroline St. N., Waterloo, ON, N2L 2Y5, Canada.\\
$^3$Argelander Institute for Astronomy, University of Bonn, Auf dem H{\"u}gel 71, 53121 Bonn, Germany.\\
$^4$Scottish Universities Physics Alliance, Institute for Astronomy, University of Edinburgh, Royal Observatory, Blackford Hill, \\Edinburgh, EH9 3HJ, UK. \\ 
$^5$Department of Physics and Astronomy, University of British Columbia, 6224 Agricultural Road, Vancouver, V6T 1Z1, BC, Canada.\\
$^6$Leiden Observatory, Leiden University, Niels Bohrweg 2, 2333 CA Leiden, The Netherlands.\\
$^7$Department of Physics and Astronomy, University of Victoria, Victoria, BC V8P 5C2, Canada.\\
$^8$Institut d'Astrophysique de Paris, UniversitÃ© Pierre et Marie Curie - Paris 6, 98 bis Boulevard Arago, F-75014 Paris, France.\\
$^9$Department of Physics, Oxford University, Keble Road, Oxford OX1 3RH, UK.\\ 
$^{10}$Institut de Ciencies de l’Espai, CSIC/IEEC, F. de Ciencies, Torre C5 par-2, Barcelona 08193, Spain.\\
$^{11}$Institute of Astronomy and Astrophysics, Academia Sinica, P.O. Box 23-141, Taipei 10617, Taiwan.\\
$^{12}$Key Lab for Astrophysics, Shanghai Normal University, 100 Guilin Road, 200234, Shanghai, China. \\
$^{13}$Kavli Institute of Particle Astrophysics and Cosmology (KIPAC), Stanford University, 452 Lomita Mall, Stanford, CA 94305, and \\
SLAC National Accelerator Laboratory, 2575 Sand Hill Road, M/S 29, Menlo Park, CA 94025, United States of America.\\
$^{14}$Max-Planck-Institut f{\"u}r Astrophysik,Karl-Schwarzschild-Stra{\ss}e 1, D-85748, Garching, Germany \\
$^{15}$Department of Physics and Astronomy, University College London, Gower Street, London WC1E 6BT, U.K.\\
$^{16}$California Institute of Technology, 1200 E California Boulevard, Pasadena CA 91125, USA.\\}

\begin{document}

\date{??}

\pagerange{\pageref{firstpage}--\pageref{lastpage}} \pubyear{2013}

\maketitle

\label{firstpage}

\begin{abstract}
We use weak gravitational lensing to analyse the dark matter halos around satellite galaxies in galaxy groups in the CFHTLenS dataset. This dataset is derived from the CFHTLS-Wide survey, and encompasses 154 deg$^2$ of high-quality shape data. Using the photometric redshifts, we divide the sample of lens galaxies with stellar masses in the range $10^{9}\msun$ to $10^{10.5} \msun$ into those likely to lie in high-density environments (HDE) and those likely to lie in low-density environments (LDE). Through comparison with galaxy catalogues extracted from the Millennium Simulation, we show that the sample of HDE galaxies should primarily ($\sim 61$\%) consist of satellite galaxies in groups, while the sample of LDE galaxies should consist of mostly ($\sim 87$\%) non-satellite (field and central) galaxies. Comparing the lensing signals around samples of HDE and LDE galaxies matched in stellar mass, the lensing signal around HDE galaxies clearly shows a positive contribution from their host groups on their lensing signals at radii of $\sim 500$--$1000$~kpc, the typical separation between satellites and group centres. More importantly, the subhalos of HDE galaxies are less massive than those around LDE galaxies by a factor $0.65\pm0.12$, significant at the 2.9$\sigma$ level. A natural explanation is that the halos of satellite galaxies are stripped through tidal effects in the group environment. Our results are consistent with a typical tidal truncation radius of $\sim 40$~kpc.
\end{abstract}


\section{Introduction}

\label{sec:intro}
In the standard picture of hierarchical structure formation, larger dark matter halos are built up through the accretion, stripping, and mergers of smaller halos. At the extremes of the halo mass spectrum, namely isolated field galaxies and galaxy clusters, we have a relatively good picture of how the mass within these structures is organized. For isolated galaxies, most of the mass is contained within a halo of dark matter, as confirmed by galaxy-galaxy weak gravitational lensing measurements \citep{BraBlaSma96, HudGwyDah98, GuzSel02, HoeFraKui03, HoeYeeGla04, ManSelKau06, HoeSch12,VelUitHoe12}. Simulations have shown that the shape of this halo can be well-approximated by an NFW density profile \citep{NavFreWhi97}, which has been confirmed observationally \citep{KleSchErb03,HoeYeeGla04,ManSelHir08}. In galaxy clusters, most of the mass also seems to lie within an NFW dark matter halo, with the constituent galaxies contributing only small perturbations to the density profile \citep{ManSelCoo06,ManSelHir08}. Gravitational lensing measurements have shown that the halos around individual galaxies within clusters are significantly smaller than the halos around comparably-luminous field galaxies, and this effect is more extreme with galaxies closer to the centres of clusters \citep{LimKneBar07,NatKneSma09}.

However, between the extremes of field galaxies and rich clusters, the picture is less clear. Since multiple galaxies must merge together to eventually form clusters, at some point the mass in the galaxies' individual halos must migrate into a shared halo. This process most likely occurs through tidal stripping: when two galaxies pass near each other, the particles in the halo of the less massive galaxy will tend to be ``stripped'' from it and thus join the more massive galaxy's halo. This effect has been demonstrated in various N-body dark matter simulations \citep{HayNavPow04,KazMayMas04,SprWanVog08}. Tidal stripping is also expected to remove hot gas from less massive galaxies, which will have the effect of cutting off their supply of cold gas and quenching their star formation in a process known as ``strangulation'' \citep{TinLar79,BalMor00}. Galaxies in dense environments are known to be significantly redder on average than field galaxies \citep{Dre80,ButOem84,MooKatLak96,BalMorYee99,BalBalNic04}, and tidal stripping may contribute to the quenching of star formation, so there is a strong motivation to understand the mechanics and timing of tidal stripping \citep{AquYan08,KawMul08}. It remains unclear, however, whether this process is rapid or gradual. This question can in part be investigated through analysis of the group and cluster scales on which tidal stripping can be observed to occur.

In this paper, we focus on galaxy groups, an intermediate mass scale between field galaxies and clusters (typically structures in the mass range $10^{12}$~$\msun \lesssim M\sbr{halo} \lesssim 10^{14}$~$\msun$ are considered groups, and more massive structures are considered clusters). Weak gravitational lensing provides the only practical tool to measure the density profiles and masses of dark matter halos around satellite galaxies within groups. Lensing analyses of groups \citep{HoeFraKui01, ParHudCar05, ManSelCoo06, JohSheWec07, HamMiyKas09, LeaFinKne10, ForHilVan12} have shown that the group lensing signal can be measured and is on average consistent with an NFW density profile. However, it is unclear how much of this signal results from a central halo, and how much is due to the contributions of satellites \citep{GilHudHil12}. As such, it is necessary to measure the lensing signals around satellites themselves to get a full picture of the mass distribution. Only limited work has been done in the group regime to date. For example, \citet{SuyHal10} studied a strong-lensing system and determined that tidal stripping did seem to occur around the satellite studied, which lies in a group of mass on the order of $10^{12} \msun$. While this result is promising, a broader base of data will be needed to develop a general understanding of the dark matter properties of satellite galaxies in galaxy groups.

In principle, it is possible to study tidal stripping using spectroscopically-derived group catalogues \citep{PasHilHar11,GilHudHil12}. Spectroscopic data allow one to more accurately identify whether a given galaxy is a ``central'' or a ``satellite'. By using velocity information in addition to projected separation, it is also possible to assess statistically whether a satellite is falling in for the first time, or has passed pericentre and hence is tidally stripped \citep{OmaHudBeh13}. However, such analyses require a large galaxy sample with both spectra and deep imaging data, and such data are expensive to acquire. However, photometric redshifts are often available alongside imaging data. Due to their large uncertainties, photometric redshifts have the drawback that groups are difficult to detect, and group-central galaxies are very difficult to identify.  This then calls for a statistical approach, calibrated by simulations, to simultaneously fit both the satellites' and groups' contributions to the stacked lensing signal. This is the approach we will take in this paper.

A similar approach was taken by \citet{ManSelKau06}, who investigated a large selection of galaxies from the SDSS, selected by environment. They tested for the presence of stripping by fitting models for the lensing signal to the galaxies in high-density environments (HDE) and low-density environments (LDE). The authors found no significant evidence of tidal stripping, but did not completely rule it out, either. This was also later attempted by \citet{HoeVel11}, who used the overlap between RCS2 and SDSS, but they were similarly unable to get a clear detection of tidal stripping. Here we will use data from the CFHTLenS collaboration, which is significantly deeper than the SDSS, and hence should provide a stronger lensing signal. We also apply a new environment estimator, which is tuned to work for photometric redshifts (see \secref{P3}) and a modified halo model designed to work with this environment estimator (see \secref{models}).

In \secref{data} of this paper, we discuss the datasets used in the  analysis and the algorithm for estimating galaxy environments. In \secref{models}, we detail the models for the lensing signals and the procedure used to fit the models to the measured signals. In \secref{results}, we present the results of the analysis and discuss possible sources of error. We conclude in \secref{conc}.

For consistency with the Millenium Simulation, we use the following cosmological parameters: $H\sbr{0} = 73$ km s$^{-1}$ Mpc$^{-1}$, $\Omega\sbr{m} = 0.25$, $\Omega\sbr{\Lambda} = 0.75$, and $\Omega\sbr{b}=0.045$. All stated magnitudes are in the AB system. Since there is no clear division between galaxy groups and galaxy clusters, we use the terminology ``galaxy groups'' throughout this paper, even though some of the structures we refer to as such would be more commonly deemed clusters.  When masses are quoted in this paper, $M$ is used to refer to the total (halo + stellar) mass of a galaxy or group, and $m$ is used to refer to the stellar mass of a galaxy, unless otherwise specified. When radial measurements are used in this paper, $R$ refers to a projected, 2D proper distance. All masses are in units of $\msun$ unless otherwise specified.

\section{Data and Simulations}

\label{sec:data}

In this section, we discuss the datasets used and how galaxies are selected for the HDE and LDE samples.  In \secref{obsdata}, we discuss the CFHTLenS survey, from which we draw our data. In \secref{P3}, we discuss the algorithm used to estimate the local density around galaxies in the sample. \secref{matching} describes how the sample is divided into matched high-density and low-density subsamples, and presents the statistics of the galaxies in the HDE and LDE samples. In \secref{simdata}, we discuss the simulations we have run to test our methods and the statistics of the galaxy samples within these simulations.

\subsection{Observations}
\label{sec:obsdata}

CFHTLenS is a 154 deg$^2$ survey (125 deg$^2$ after masking) \citep{ErbHilMil12}, based on the Wide component of the Canada-France-Hawaii Telescope Legacy Survey \citep{HeyVanMil12}, which was observed in the period from March 22nd, 2003 to November 1st, 2008, using the MegaCam instrument \citep{BouChaAbb03}. It consists of deep, sub-arcsecond, optical data in the $u^{*}g'r'i'z'$ filters. CFHTLS-Wide observations were carried out in four high-galactic-latitude patches:

\begin{itemize}
\item W1: 72 pointings; RA=$02\spr{h}18\spr{m}00\spr{s}$, Dec=$-07\spr{d}00\spr{m}00\spr{s}$
\item W2: 25 pointings; RA=$08\spr{h}54\spr{m}00\spr{s}$, Dec=$-04\spr{d}15\spr{m}00\spr{s}$
\item W3: 49 pointings; RA=$14\spr{h}17\spr{m}54\spr{s}$, Dec=$+54\spr{d}30\spr{m}31\spr{s}$
\item W4: 25 pointings; RA=$22\spr{h}13\spr{m}18\spr{s}$, Dec=$+01\spr{d}19\spr{m}00\spr{s}$.
\end{itemize}

Shapes have been measured with the LensFit shape measurement algorithm for galaxies with $i' < 24.7$ \citep{MilHeyKit12}, giving an effective galaxy density of 11 sources/arcmin$^2$ in the redshift range $0.2 < z\sbr{phot} < 1.3$ \citep{HeyVanMil12}. Photometric redshifts are available for the entire survey, with a typical redshift uncertainty of  $\sim0.04(1+z)$ \citep{HilErbKui12}. We use all fields in the survey, not simply those that passed the systematics tests for cosmic shear measurements \citep{HeyVanMil12}. It has been demonstrated that fields with systematics that may affect cosmic shear have no effect galaxy-galaxy lensing measurements \citep{VelUitHoe12}, and the analysis in this paper requires as many lens-source pairs as possible.

We use the stellar mass estimates described by \citet{VelUitHoe12}, obtained by fitting spectral energy distribution (SED) templates, following the method of \citet{IlbSalLe10}. These stellar masses were found to be in rough agreement with deeper data such as WIRDS, which includes NIR filters \citep{BieHudMcC12}, up to $z = 0.8$. 

Since we perform a differential measurement between samples, an overall bias in the stellar masses would not affect our results. It is possible, however, for a relative bias in the stellar mass estimates of red and blue galaxies to impact our results. This possibility is investigated further in \secref{bias}.

For this paper, we use take all unmasked galaxies with photometric redshifts in the range $0.2 < z\sbr{phot} < 0.8$ as lens candidates. We divide these into HDE and LDE samples as described next.

\subsection{Determining Environment: The P3 Algorithm}
\label{sec:P3}

It is not a trivial matter to determine which galaxies are members of groups. Even when spectroscopic redshifts are available, the peculiar velocities of galaxies make it impossible to determine the memberships of groups with absolute certainty \citep{RobNorDri11}. When only photometric redshifts are available, the best we can do is to select galaxies that are likely to be members of groups. To do so, we use a modified version of the Photo-z Probability Peaks (P3) algorithm \citep{GilHud11}. The P3 algorithm generates a 3-D density field by smoothing the distribution of galaxies in the redshift direction according to the probability distribution function of their photometric redshifts. The algorithm identifies peaks in the pseudo-three-dimensional field with group centres. Here we do not use the group centres, but rather use the entire P3 density field to identify overdense regions. Rather than use the local P3 overdensity itself, we restrict ourselves to regions in which we have high confidence in the overdensity, and instead use the signal-to-noise (S/N) of the local overdensity, under the assumption that galaxies in overdense regions of space are more likely to be in groups than galaxies in underdense regions.

We now briefly review the technical details of the P3 algorithm. To determine the S/N of a given test galaxy, the P3 algorithm compares the density of galaxies within a circular aperture ($R = 0.5$~Mpc) surrounding each test galaxy to the density of galaxies within a larger annulus ($R\sbr{inner} = 1$~Mpc, $R\sbr{outer}=3$~Mpc) surrounding each test galaxy (to approximate the background density). The contribution of each galaxy to this measurement is weighted by the probability that this galaxy lies at the same redshift as the test galaxy (by taking the integral of the photo-z probability distribution function over a thin redshift slice). This gives the overdensity:
\begin{equation}
\delta = \frac{\rho\sbr{ap}-\rho\sbr{annu}}{\rho\sbr{annu}}\textrm{,}
\end{equation}
where $\rho\sbr{ap}$ and $\rho\sbr{annu}$ are the weighted densities of galaxies within the aperture and annulus surrounding the test galaxy, respectively. This value can take the range $-1 < \delta < \infty$, which negative values corresponding to regions less dense than the background density, and positive values to overdense regions. We then estimate the noise in this value by assuming a Poisson distribution for galaxies:

\begin{equation}
\sigma\sbr{Poisson} = \sqrt{\left(\frac{\rho\sbr{ap}}{n\sbr{ap}}\right)^{2}+\left(\frac{\rho\sbr{annu}}{n\sbr{annu}}\right)^{2}}\textrm{,}
\end{equation}%
where $n\sbr{ap}$ and $n\sbr{annu}$ are the numbers of galaxies in the aperture and annulus respectively with more than a threshold weight\footnote{We use a threshold weight here of a $>0.001\%$ chance of lying within a redshift of 0.01 of the test galaxy.}. From this, we calculate the S/N $\equiv \delta/\sigma\sbr{Poisson}$ for each test galaxy. Note that this S/N can take negative values, assuming $\delta$ is negative. The distribution of galaxies' S/Ns that results from this calculation depends on the choice of threshold weight used, so our choices of S/N limits are not universally applicable. We picked limits of S/N $>$ 2 for the high-density sample and S/N $<$ 0 for the low-density sample based on an analysis of the simulated galaxy catalogues to maximize the expected signal for tidal stripping.\footnote{In a rough approximation, the expected signal-to-noise of a stripping measurement is proportional to $(f\sbr{sat,HDE}-f\sbr{sat,LDE})\sqrt{N\sbr{HDE}^{-1}+N\sbr{LDE}^{-1}}$, where $f\sbr{sat,HDE}$ and $f\sbr{sat,LDE}$ are the fractions of satellites in the HDE and LDE samples respectively, and $N\sbr{HDE}$ and $N\sbr{LDE}$ are the number counts of galaxies in the HDE and LDE samples respectively. We calculated this value for various S/N cuts, and the combination of S/N $>$ 2 for the HDE sample and S/N $<$ 0 for the LDE sample provided the best expected signal-to-noise for a stripping measurement.}

Since this environment estimator provides us with galaxy samples biased to lie in high- and low-density environments, we cannot use the standard halo model \citep[eg.][]{ManSelKau06,VelUitHoe12} for fitting our lensing signals. Instead, the models we use are calibrated from simulations and are detailed in \secref{models}.

\subsection{Galaxy Matching}
\label{sec:matching}
\subsubsection{Matching Algorithm}

We use the S/N values obtained for each galaxies in \secref{P3} to form two samples of galaxies from the catalogues. As we cannot ensure that a pair of random samples of galaxies in high- and low-density environments will have the same distribution of stellar mass and redshift as each other, we perform a matching between galaxies with S/N $>$ 2 and galaxies with S/N $<$ 0 as follows:

\begin{enumerate}
\item For each galaxy with S/N $>$ 2, we search through all galaxies with S/N $<$ 0 within the same pointing\footnote{Matching only within the same pointing is done to conserve computational time.}.
\item For each S/N $<$ 0 galaxy, if its stellar mass differs from the stellar mass of the S/N $>$ 0 galaxy by more than $20\%$ of the latter's mass, we exclude it as a possible match.
\item For each remaining S/N $<$ 0 galaxy, we calculate a quality-of-match value:
\begin{equation}
d = \sqrt{\left(\frac{z\sbr{H}-z\sbr{L}}{z\sbr{H}}\right)^2+\left(10(\log{m\sbr{H}}-\log{m\sbr{L}})\right)^2}
\end{equation}%
where $z\sbr{H}$ and $z\sbr{L}$ are the redshifts of the S/N $>$ 2 and S/N $<$ 0 galaxies, respectively, and $m\sbr{H}$ and $m\sbr{L}$ are their stellar masses. This form significantly prioritizes a match in mass over redshift.
\item We select the four S/N $<$ 0 galaxies with the lowest $d$ values as matches for this S/N $>$ 2 galaxy. If there are fewer than four match candidates, we assign them all as matches.
\item Assuming at least one match was found for it, we add this S/N $>$ 2 galaxy to the HDE sample, and we set its weight equal the number of matches we found. (This weight is later applied when we stack lensing signals together, and this modification is necessary to ensure the mass distributions of the HDE and LDE samples are comparable.)
\item We assign all match galaxies to the LDE sample. If they were not already in the LDE sample, we set each of their weights to 1. Otherwise, we increase their weights by 1.
\end{enumerate}

The resultant mass and redshift distributions of this scheme are assessed in \secref{statobs} and \secref{statsim}. 

\subsubsection{Statistics of Galaxy Selection}
\label{sec:statobs}

\begin{figure*}
\centering
\includegraphics[scale=0.4]{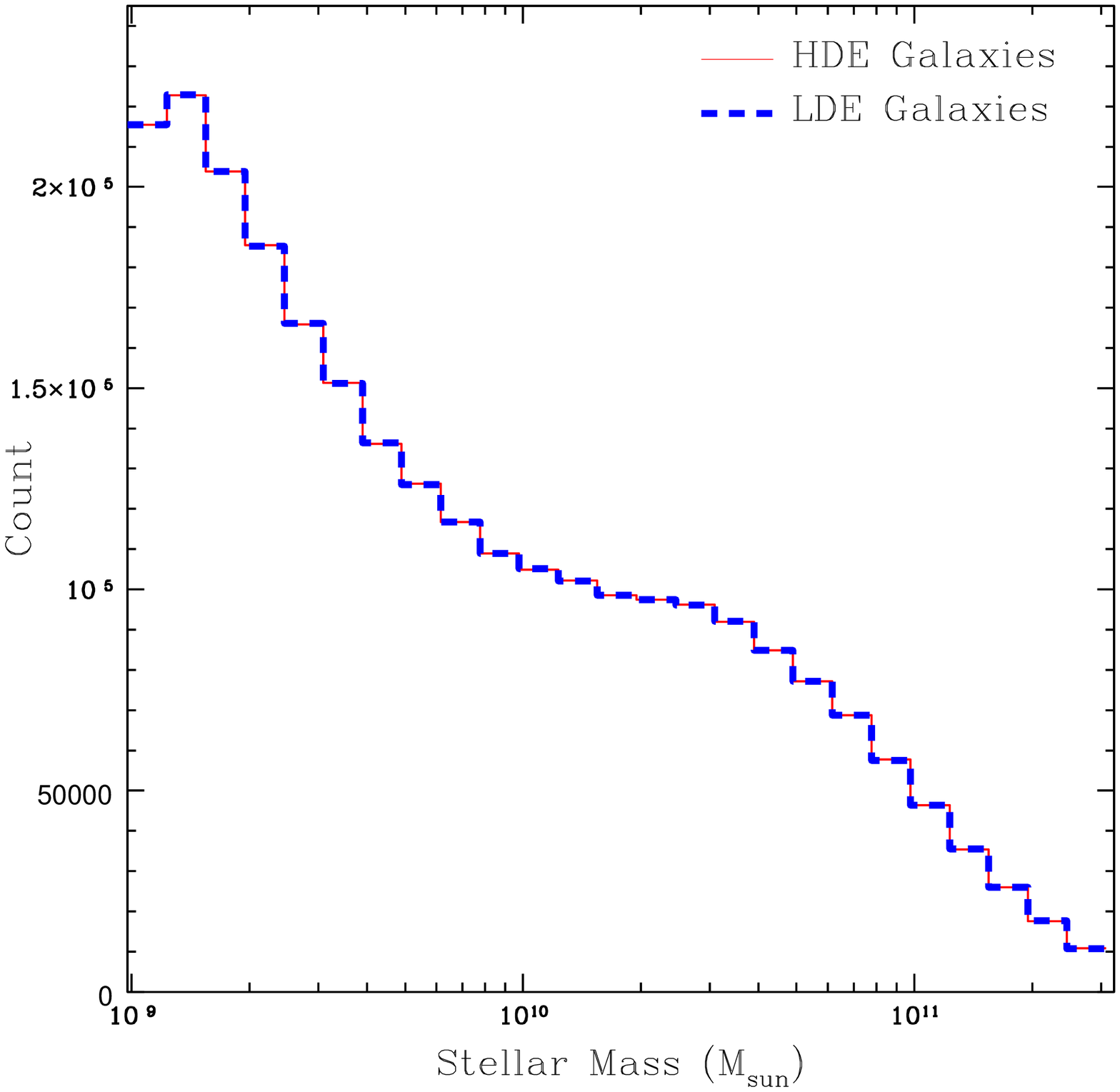}
\includegraphics[scale=0.4]{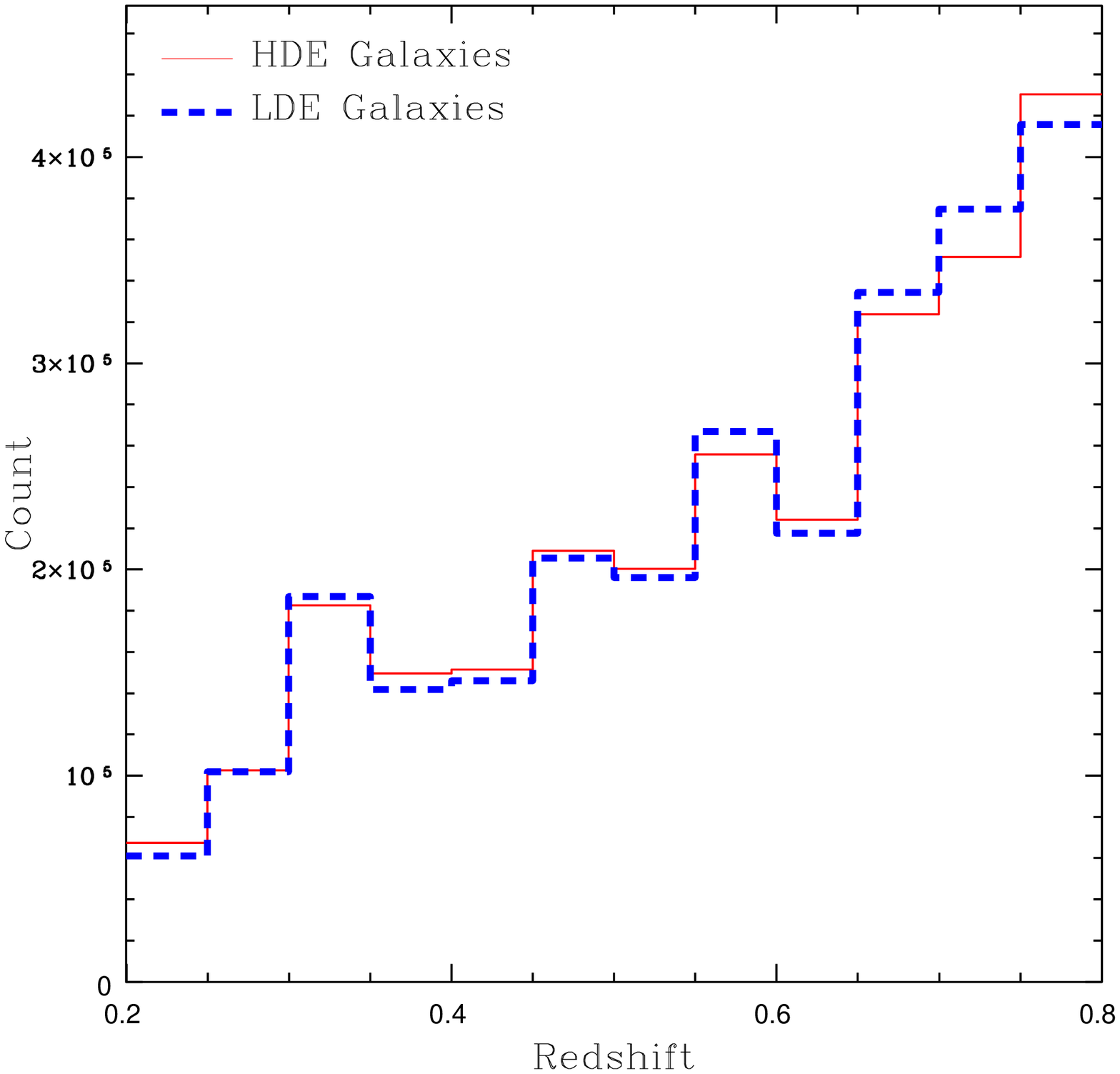}
\caption[]{The distributions of stellar mass (left panel) and redshift (right panel) for the samples of HDE (solid line) and LDE (dashed line) galaxies, which, because of our matching algorithm, are virtually identical. The redshift distributions differ slightly between HDE and LDE galaxies, but there is no apparent trend to the deviation.}
\label{fig:galaxyhists}  
\end{figure*}

\figref{galaxyhists} shows the distributions of stellar mass and redshift for the HDE and LDE samples of lens galaxies in the CFHTLenS. The matching scheme results in a nearly identical distribution of stellar masses for HDE and LDE galaxies, and a very similar distribution of redshifts.

\begin{table}
\centering
\begin{tabular}{| l l l l l l l l l l l l l|}
 & & & \multicolumn{2}{|c|}{HDE} & & \multicolumn{2}{c|}{LDE} \\
\hline
\phn\phn $\log{m}$ & \phn$\overline{z}$ & & $f\sbr{red}$ & $f\sbr{blue}$ & & $f\sbr{red}$ & $f\sbr{blue}$ \\ \hline
\phd\phn\phn 9--9.5 & 0.57 & & 0.13 & 0.73 & & 0.08 & 0.80 \\
\phn 9.5--10 & 0.56 & & 0.28 & 0.60 & & 0.18 & 0.70 \\
\phd\phn 10--10.5 & 0.56 & & 0.54 & 0.30 & & 0.44 & 0.38 \\
10.5--11 & 0.57 & & 0.78 & 0.10 & & 0.72 & 0.13 \\
\phd\phn 11--11.5 & 0.57 & & 0.95 & 0.02 & & 0.90 & 0.03 \\ \hline
\phd\phn\phn 9--10.5 & 0.56 & & 0.43 & 0.43 & & 0.33 & 0.51 \\
\hline
\end{tabular}
\caption{Statistics of galaxies in various stellar mass bins in the CFHTLenS survey, as a function of environment. $\overline{z}$ is the mean redshift of the bin. $f\sbr{red}$ is the fraction of galaxies that are red, and $f\sbr{blue}$ is the fraction that are blue, determined by the best-fit photometric templates and defined in the same manner as by \citet{VelUitHoe12}. Fractions do not add to unity as not all galaxies are classified as ``red'' or ``blue.'' See \citet{VelUitHoe12} for further explanation. All average values and fractions assume galaxies are weighted by their stellar masses.}
\label{tab:obsstats}
\end{table}

\tabref{obsstats} shows statistics for lens galaxies in the HDE and LDE samples in the CFHTLenS, for various stellar mass bins. The HDE sample contains a higher fraction of red galaxies than the LDE sample, as expected, but the difference is at most $10\%$ for a given stellar mass bin. This difference in the fractions of red and blue galaxies could in principle lead to a spurious detection of stripping if there is a relative bias in the stellar mass estimates between red and blue galaxies. This issue is discussed further in \secref{bias}.

\subsubsection{Measuring the Lensing Signal}

\begin{figure*}
\centering
\includegraphics[scale=0.4]{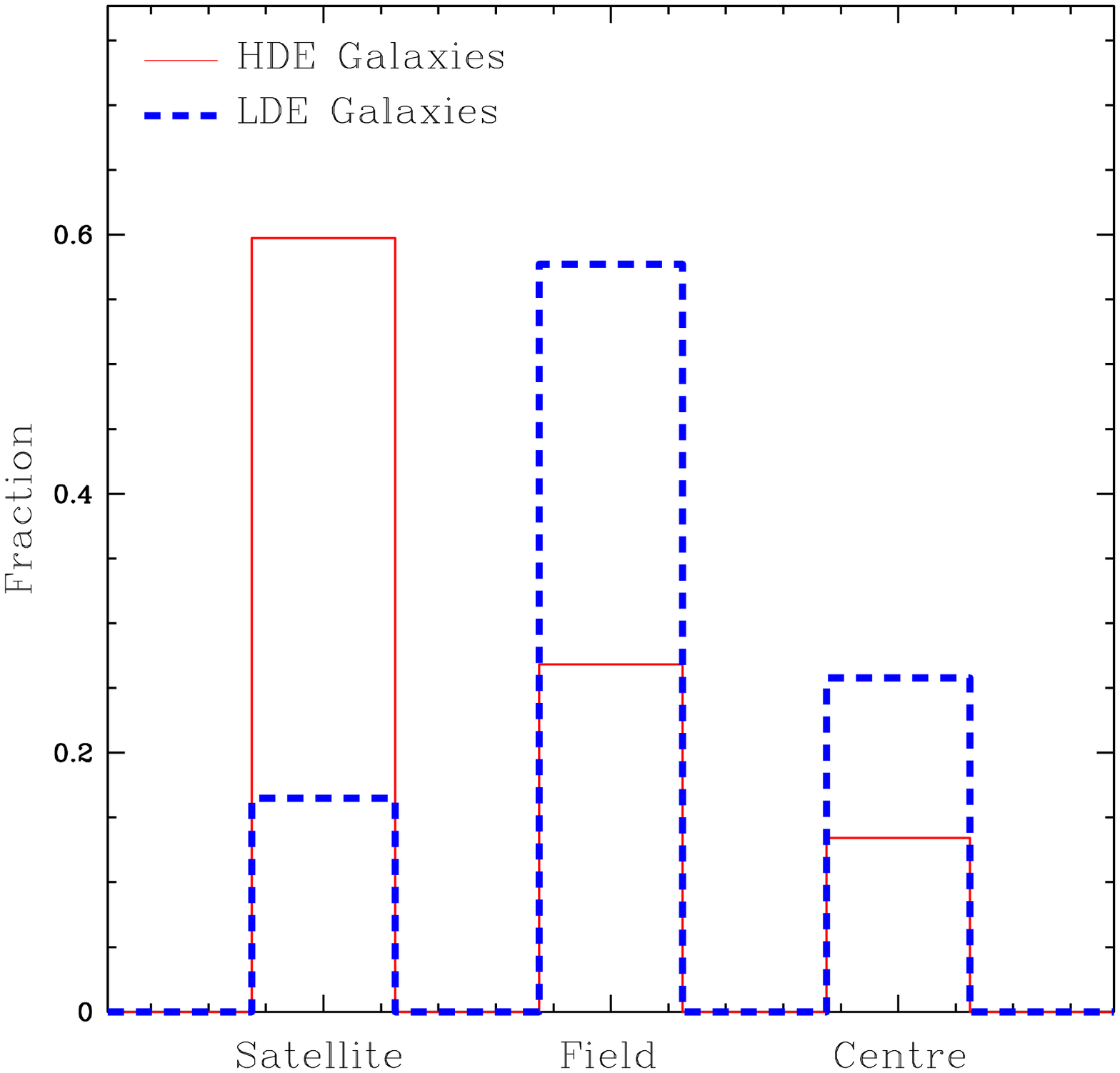}
\includegraphics[scale=0.4]{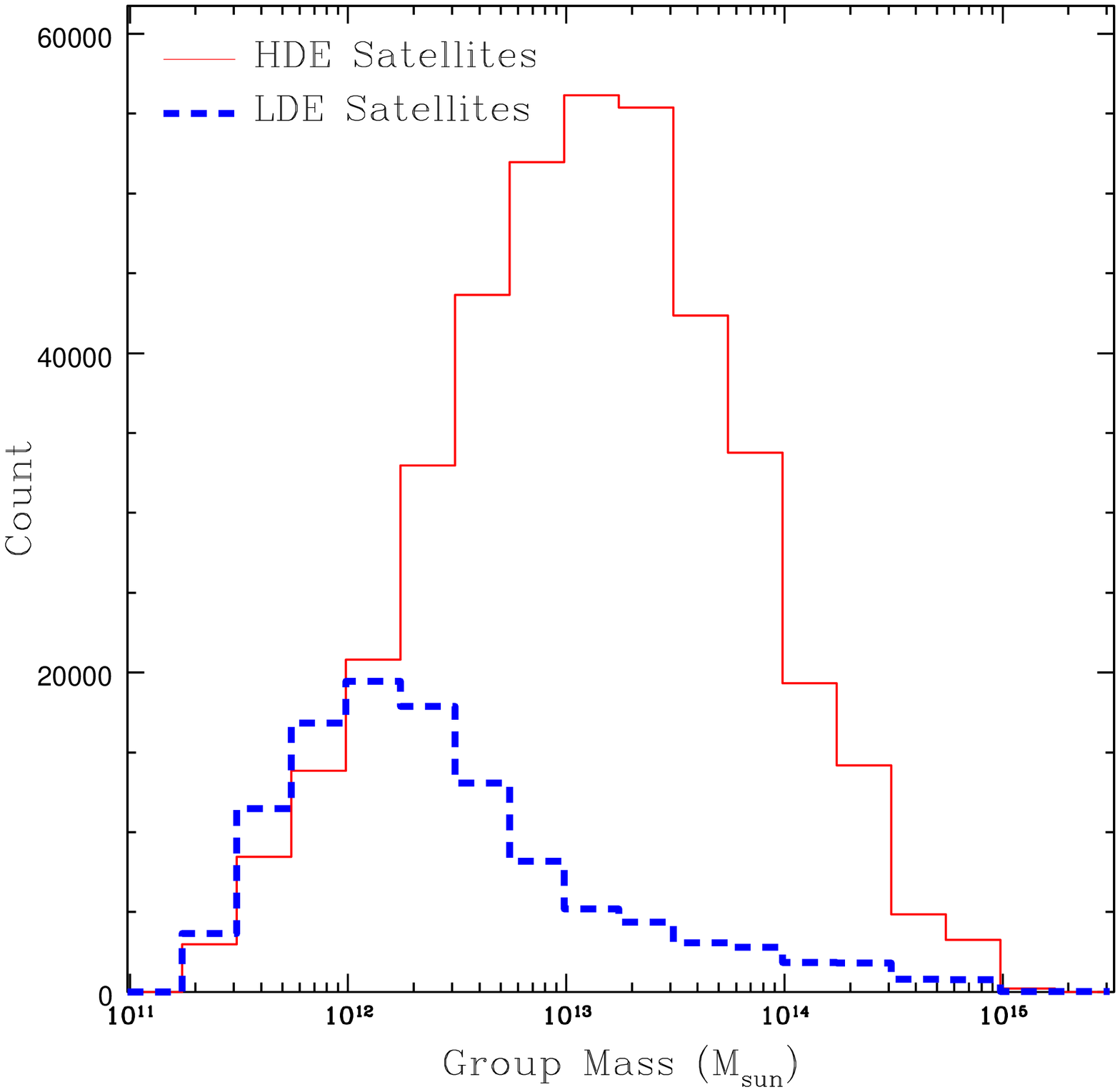}
\caption[]{The distributions of the types of galaxies classified as HDE and LDE in the simulations (left), and, of those classified as satellites, the distributions of the masses of the groups in which they reside (right).}
\label{fig:simgalhist}  
\end{figure*}

\begin{table*}
\centering
\begin{tabular}{| l l l l l l l l l l l l l|}
 & & & & \multicolumn{4}{|c|}{HDE} & & \multicolumn{4}{c|}{LDE} \\
\hline
\phn\phn $\log{m}$ & $\overline{M}$ & $\overline{z}$ & & $f\sbr{sat}$ & $f\sbr{field}$ & $f\sbr{cen}$ & $\overline{M\sbr{host}}$ & & $f\sbr{sat}$ & $f\sbr{field}$ & $f\sbr{cen}$ & $\overline{M\sbr{host}}$  \\ \hline
\phd\phn\phn 9--9.5 & \phn17 & 0.37 & & 0.54 & 0.42 & 0.05 & 4700 & & 0.14 & 0.79 & 0.07 & 1635 \\
\phn 9.5--10 & \phn32 & 0.45 & & 0.62 & 0.25 & 0.13 & 4000 & & 0.15 & 0.61 & 0.24 & 1500 \\
\phd\phn 10--10.5 & \phn80 & 0.51 & & 0.64 & 0.10 & 0.26 & 4300 & & 0.13 & 0.32 & 0.54 & 1900 \\
10.5--11 & 390 & 0.50 & & 0.45 & 0.02 & 0.53 & 5600 & & 0.09 & 0.12 & 0.79 & 3400 \\ \hline
\phn 9--10.5 & \phn63 & 0.48 & & 0.63 & 0.16 & 0.21 & 4300 & & 0.14 & 0.43 & 0.43 & 1800 \\
\hline
\end{tabular}
\caption{Statistics of galaxies in the Millennium simulation for various stellar mass bins, using our models for estimating halo mass and environment. $\log{m}$ is the stellar mass bin. $\overline{M}$ is the mean halo mass of the galaxies in this bin in units of $10^{10} \msun$, and $\overline{z}$ is their mean redshift. $f\sbr{sat}$, $f\sbr{field}$, and $f\sbr{cen}$ are the fractions of galaxies that are satellites, field galaxies, and group centrals, respectively. $\overline{M\sbr{host}}$ is the mean mass of the host group for satellite galaxies in units of $10^{10} \msun$. All values assume galaxies are weighted by their stellar masses.}
\label{tab:simstats}
\end{table*}

To calculate the lensing signal around the HDE and LDE lens galaxies, we stack together all galaxies in a particular sample and stellar mass bin\footnote{This process is performed one pointing at a time due to computational limitations, and all pointings are stacked together in the end.}. We then bin all lens-source pairs (only using pairs where $z\sbr{phot,source} > z\sbr{phot,lens}+0.1$) based on the projected distance between the lens and source, calculated at the redshift of the lens. For each pair, we calculate the tangential ellipticity of the source relative to the lens, $g_t$, and convert this into units of surface mass density gradient:
\begin{equation}
\Delta\Sigma = \Sigma\sbr{crit}g_t\textrm{,}
\end{equation}%
where
\begin{equation}
\Sigma\sbr{crit} = \frac{c^2 D\sbr{s}}{4\pi G D\sbr{ls}D\sbr{l}}\textrm{,}
\end{equation}
and $D\sbr{s}$ is the angular diameter distance to the source, $D\sbr{l}$ is the angular diameter distance to the lens, and $D\sbr{ls}$ is the angular diameter distance from the lens to the source. This measurement relates to the projected mass of the lens through \citep{ManHirSel05}:
\begin{equation}
\left<\Delta\Sigma(R)\right> = \overline{\Sigma(<R)}-\overline{\Sigma(R)}\textrm{,}
\label{eq:mainlens}
\end{equation}
where $\overline{\Sigma(<R})$ is the surface density averaged for all points contained within radius $R$, and $\overline{\Sigma(R)}$ is the average surface density at radius $R$. This prescription works even for mass distributions that are not axisymmetric, as long as all points in a given annulus around a lens object are stacked. We compute the error in this value empirically from the scatter in calculated $\Delta\Sigma$ values for each lens-source pair.

For the calculations of error in our model fits, we assume the noise in all radial annuli is independent. Strictly speaking, this isn't true, as there is a small correlation between the ellipticities of nearby sources, but this effect is negligible except at extremely large radial annuli. For computational simplicity, we do not apply the $c2$ correction\footnote{The $c2$ correction is an empirical correction to the $e2$ component of source ellipticity, based on the assumption that the mean e2 across a given field should be close to zero.} to source ellipticities in our analysis. Because galaxy-galaxy measurements stack lens-source pairs over all position angles, they are insensitive to this correction\citep[see][for further explanation and justification of this]{VelUitHoe12}. Moreover, here we are interested in a differential measurement between galaxy-galaxy lensing samples, and so we expect our results to be highly robust to this effect. 

\subsection{Simulated Galaxy Catalogues}
\label{sec:simdata}

We require simulations in order to calibrate the fractions of satellite and central galaxies in our samples, and to test our methods for modelling the lensing signals around galaxies. The simulations are based on the semi-analytic models of \citet{DeBla07} which are based on the Millenium Simulation \citep{SprWhiJen05,LemVir06}. We use for our analysis a set of thirty-two 16~deg$^2$ ``lightcone'' fields by \citet{HilHarWhi09}. We assign photo-z errors consistent with those in CFHTLenS and apply the P3 algorithm to the simulated data. This allows us to select galaxies in the same manner as is done with the CFHTLenS dataset.

\subsubsection{Statistics of Simulated Catalogues}
\label{sec:statsim}

\figref{simgalhist} shows the distributions of galaxy types for the mock HDE and LDE samples drawn from the Millennium Simulation, and, for the satellite galaxies within each sample, the distribution of the masses of the groups in which they reside. We classify galaxies as ``central'' (the most massive in a group), ``satellite'' (in a group but not the most massive) or ``field'' (not in a group).  \tabref{simstats} shows the distributions of galaxy types for the HDE and LDE samples for various stellar mass bins. This shows that the fraction of satellites in the HDE sample remains roughly constant with stellar mass, and decreases slightly with stellar mass in the LDE sample. For both samples, the fraction of centrals rises with stellar mass, while the fraction of field galaxies falls. Neither sample shows any significant change with stellar mass in the mean mass of the host groups for satellites, except for a rise in the most massive stellar mass bin tested. HDE satellites are observed to reside in groups of $\sim 4\times 10^{13}$ \msun. In contrast, for the small fraction of LDE galaxies that are satellites, the characteristic host halo mass is $\sim 1.8\times 10^{13}$ \msun.

\subsubsection{Simulated Lensing Signals}
\label{sec:simlens}

In order to simulate lensing signals for the Millennium Simulation catalogues, we use the same methods as in \citet{GilHudHil12}. In short, we assume all galaxies and group centres are surrounded by spherical truncated NFW halos \citep{Bar96,HamMiyKas09}, using the model of \citet{BalMarOgu09}, and estimating halo mass from stellar mass by using equation 3 from \citet{GuoWhiLi10} (to better match the stellar-dark mass ratio in the CFHTLenS):

\begin{eqnarray}
\label{eq:MhfromMs}
0.129\times\frac{m\sbr{halo}}{m\sbr{stellar}} = \Biggl(\left(\frac{m\sbr{halo}}{10^{11.4} \textrm{M}_{\odot}}\right)^{-0.926}\nonumber \\
+\left(\frac{m\sbr{halo}}{10^{11.4} \textrm{M}_{\odot}}\right)^{0.261}\Biggr)^{2.44}\textrm{.}
\end{eqnarray}

We form two versions of the simulated shear catalogue, one in which we simulate the effects of stripping by decreasing the truncation radii of satellite galaxies' halos and allocating the lost mass to group centres' halos (``Stripping'') and one in which we do not (``No Stripping'').  The amount of stripping is assumed to depend on distance from the centre of the host halo and is given by equation (5) of \citet{GilHudHil12}, which is based on data in figure 15 of \citet{GaoWhiJen04}.  With this prescription, we find that the mean retained mass after stripping is approximately $40\%$ of the initial mass. We then calculate shapes for all background galaxies by assuming initially zero ellipticity in both components, then applying shear due to each nearby halo between the source galaxy and the observer.

\section{Models and Signal Fitting}
\label{sec:models}

\begin{figure}
\centering
\includegraphics[scale=0.4]{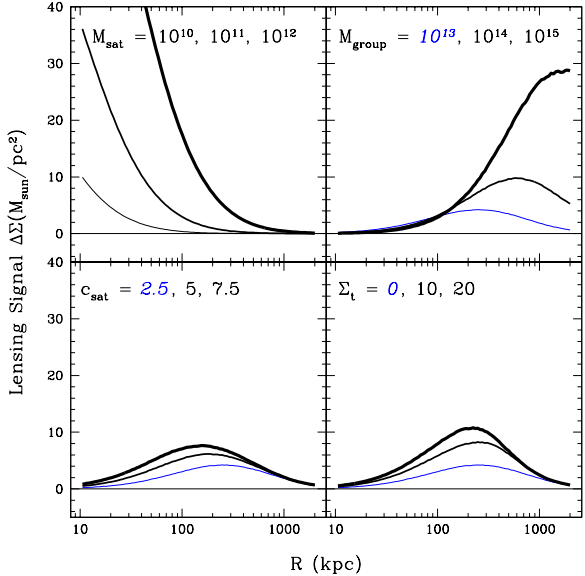}
\caption{An illustration of how the one-halo term varies with satellite mass (top left); and how the offset group halo term (\eqref{ogt}) varies with group mass (top right), satellite concentration (bottom left), and density threshold (bottom right). Plotted values of the parameters, with italicized parameter corresponding to the value used for other plots: $M\sbr{sat} = 10^{10}$, $10^{11}$, $10^{12}$~$\msun$; $M\sbr{group} = \mathit{10^{13}}$, $10^{14}$, $10^{15}$~$\msun$; c$\sbr{sat} = \mathit{2.5}$, $5$, $7.5$, $\Sigma_t = \mathit{0}$, $10$, $20$~$\msun/pc^{2}$. Increasing line weight corresponds to increasing the varying parameter. The fraction of satellites which reside in groups is not illustrated, as it is a simple scaling of the group halo term; it is fixed to $0.6$ for these plots.}
\label{fig:samples}
\end{figure}

We expect the lensing signal around galaxies in the HDE sample to be reasonably well-described by the following halo model \citep[see eg.][]{VelUitHoe12}:
\begin{equation}
\Delta\Sigma = \Delta\Sigma\sbr{1h}+f\sbr{sat}\Delta\Sigma\sbr{OG}+\Delta\Sigma\sbr{2h}
\end{equation}%
where $\Delta\Sigma\sbr{1h}$ is the ``one-halo'' term, $f\sbr{sat}$ is the fraction of galaxies in the sample that are satellites $\Delta\Sigma\sbr{OG}$ is the ``offset group halo'' term, and $\Delta\Sigma\sbr{2h}$ is the ``two-halo'' term, as described below:

\begin{enumerate}
\item One-halo term: The lensing signal that results from the galaxy's own dark matter halo.
\item Offset group halo term: This is the contribution to the lensing signal a satellite caused by the presence of its group's halos.
\item Two-halo term: Galaxies will typically reside near other massive structures, which results in a contribution to the lensing signal at large radii.
\end{enumerate}

Since galaxies in the HDE sample are more likely lie in overdense regions, we cannot apply exactly the same halo model as eg. \citet{VelUitHoe12}, who use all galaxies independent of environment. This primarily affects the offset group halo term. See \secref{ogt} below for an explanation of how we modify our halo model to account for this.

For LDE galaxies, we expect the signal to be described by the form:
\begin{equation}
\Delta\Sigma = \Delta\Sigma\sbr{1h}+\Delta\Sigma\sbr{UD}
\end{equation}%
where  $\Delta\Sigma\sbr{UD}$ is the ``underdensity'' term, which is the effective contribution from the fact that galaxies in an underdense environment will see a negative contribution to their lensing signal at large radii. This effect is analogous to the offset group halo term, except arising from an underdensity instead of an overdensity.

We can best compare the lensing signals that result from stacks of HDE and LDE galaxies by fitting the signals with a model profile, and comparing these fits. The model profile for the HDE sample includes just the ``one-halo'' and ``offset group halo'' terms. Since the ``underdensity'' and ``two-halo'' terms are only significant at relatively large radii, we can safely ignore them if we do not fit the profiles out to large radii. We discuss this further in \secref{modelinac}.

These components are discussed in the subsections below, and we discuss the procedure we use to fit a model to the data in \secref{fitpro}.

\subsection{One-halo term}

For the one-halo term, we assume that all galaxies reside in a dark matter halo that can be approximated with a truncated NFW density profile, as formulated by \citet{BalMarOgu09}. This model has three free parameters: the halo mass $M\sbr{200}$, concentration $c$, and the truncation parameter $\tau \equiv r\sbr{trunc}/r\sbr{s}$. In practice, we have found that the signal is not strong enough to simultaneously constrain all three parameters. Therefore, for simplicity, in our default fits discussed below, we fit $M\sbr{200}$, with $c$ fixed by:
\begin{equation}
\label{eq:cfm}
c = 4.67\times\left(\frac{M\sbr{200}}{10^{14} h^{-1} \msun}\right)^{-0.11}\textrm{,}
\end{equation}%
taken from \citet{NetGaoBet07}, and we also fix $\tau = 2c$, which is a reasonable value for unstripped halos \citep{HilWhi10,OguHam11}. 

In \secref{1hc}, we investigate alternative fits in which $c$ or $\tau$ are allowed to be depend on environment. 

\subsection{Offset group halo term}
\label{sec:ogt}
Since the P3 algorithm biases our galaxy selection such that the HDE sample predominantly consists of galaxies within groups, we cannot use the standard halo model \citep[eg.][]{VelUitHoe12} to calculate the contributions of nearby groups. Instead, we make the assumption that the sample consists of a fraction $f\sbr{sat}$ satellites, and the rest are either central or field galaxies. The central and field galaxies will only have a one-halo component in their lensing signals, while satellites will have both the one-halo component and a contribution from their host groups. In order to model the average contribution of group halos to the lensing signal around galaxies in the HDE sample, we assume that it takes the following form:

\begin{equation}
\Delta\Sigma\sbr{OG}(R) = \Delta\Sigma\sbr{host}(R,R_s) {\rm P}(R_s) dR_s\textrm{,}
\end{equation}%
where $R_S$ is the projected separation between a satellite and the group centre, $\Delta\Sigma\sbr{host}(R,R_s)$ is the contribution of the group halo to the lensing signal around a point at projected distance R from the group centre:

\begin{eqnarray}
\Delta\Sigma\sbr{host}(R,R_s) = \overline{\Sigma\sbr{host}(<R,R_s)} - \overline{\Sigma\sbr{host}(R,R_s)} \nonumber \\
= \frac{1}{\pi R^{2}}\int_{0}^{R} 2\pi R'\int^{2\pi}_{0} \Sigma\sbr{host}(R_g)d\theta dR' \nonumber \\
- \frac{1}{2\pi}\int^{2\pi}_{0} \Sigma\sbr{host}(R_g)d\theta\textrm{,}
\label{eq:ogt}
\end{eqnarray}%
where $\Sigma\sbr{host}(R_g)$ is the projected surface density of the host group's halo at projected radius $R_g = \sqrt{R'^{2} + R_{s}^{2} - R'R_{s}\cos{\theta}}$; and P$(R_s)$ is the probability that a satellite in the sample will reside a distance $R_s$ from its host group's centre. We assume P$(R_s)$ takes the form:

\begin{equation}
{\rm P}(R_s) = \frac{1}{M\sbr{N}} 2\pi R_s \Sigma(R_s, M\sbr{gr},c\sbr{sat}){\rm P}\sbr{HDE}(R_s)\textrm{,}
\end{equation}%
where $M\sbr{N}$ is a normalization factor, $\Sigma(R_s, M\sbr{gr}, c\sbr{sat})$ is the projected surface density of an NFW halo with mass equal to the mass of the host group, $M\sbr{gr}$, but a concentration $c\sbr{sat}$, different from the dark matter concentration $c$.  Analyses of the satellite density in groups and clusters \citep{LinMohSta04,BudKopMcC12} have indicated that the spatial distribution of satellites can be well-modelled in this way, assuming an NFW density profile with concentration $\sim 2.5$, which is lower than the typical concentration of the dark matter halo by a factor of $\sim$2.

The term P$\sbr{HDE}(R_s)$ is the probability that a satellite at a distance $R_s$ from the host group's centre will be included in the HDE sample. The form of P$\sbr{HDE}(R_s)$ is determined by the selection effects inherent in the P3 algorithm. To first order, P3 selects galaxies in regions of high projected surface density for the HDE sample. We thus model P$\sbr{HDE}(R_s)$ as a smooth cut-off based on the projected surface density of the group. We wish for it to converge to P$\sbr{HDE}(R_s) = 1$ for $\Sigma(R_s) \gg \Sigma\sbr{t}$, and converge to P$\sbr{HDE}(R_s) = 0$ for $\Sigma(R_s) \ll \Sigma\sbr{t}$, and so we choose the following functional form, which has these properties:

\begin{equation}
{\rm P}\sbr{HDE}(R_s) = \frac{\Sigma(R_s)^{2}}{\Sigma(R_s)^{2}+\Sigma\sbr{t}^{2}}\textrm{,}
\end{equation}%
where $\Sigma(R_s)$ is the projected surface density for a satellite at distance $R_s$ from a group centre and $\Sigma\sbr{t}$ is the threshold density. As we have no prior justification for any specific density threshold to use, we leave this parameter free, to be fit by our algorithm. 

For the HDE sample, we fix $f\sbr{sat}$ to the value found in the mock HDE sample from the Millennium Simulation. We do not expect this simulated result to perfectly match the fraction of satellites we might find in the CFHTLenS dataset, and we investigate what impact a different $f\sbr{sat}$ might have in \secref{modelinac}. For the LDE sample, we don't include this term, as the form of the measured lensing signal in both simulated and CFHTLenS data shows that the underdensity signal dominates at large radii.

We choose to model the offset group halo term as if all groups are of the same mass. We tested using a distribution of group masses, and the resulting signal was not appreciably different from the single-mass signal. The use of a distribution of group masses did tend to increase the resultant signal (the difference scaling with the spread of the mass distribution), even when the mean mass is fixed, and so the single-mass model will likely underestimate the mean host halo mass.

\figref{samples} illustrates how the modeled one-halo term varies with satellite halo mass, as well as how the fitted offset group halo terms varies with the group halo mass $M\sbr{group}$, satellite concentration c$\sbr{sat}$ and threshold surface density $\Sigma\sbr{t}$.

\subsection{Underdensity signal}
\label{sec:under}

Galaxies in the LDE sample are selected to lie in S/N $<$ 0 regions, which are underdense ($\delta < 0$) compared to a surrounding annulus with inner radius 1~Mpc and outer radius 3~Mpc. Similarly to how galaxies in groups have a positive contribution to their lensing signal from the offset overdensity in which they reside, galaxies in underdense regions will have a negative contribution to their lensing signal on larger scales due to the fact that their local environment is less dense than the surrounding environment. This effect has been observed in both the CFHTLenS dataset, as well as in the simulations.

The expected form of this negative lensing signal has not been well-studied, so there is no functional form which we expect it to take. We have attempted to fit this signal with the same functional form as the group halo term, multiplied by a negative free term, but this failed to provide a suitable fit to either the simulated or to the CFHTLenS data. Note in the right panel of \figref{widesig} that the minimum value for the LDE signal is at a higher projected radius than the peak of the offset group halo term.

To handle this effect, for the LDE sample, we only fit the lensing signal for $R < 400$~kpc, where the one-halo term dominates the signal. 

\subsection{Fitting Procedure}
\label{sec:fitpro}

For all fits, we use radial bins of $25$~kpc$< R < 2000$~kpc. We tested constraining the fits to a lower maximum radius, and this had no noticeable effect on the fitted satellite halo masses. Fitting to a lower maximum radius only altered the fitted group mass, making it less well-constrained.

We use a two-step procedure to fit the models to the lensing signals. Because our models are relatively simple, they are not perfect fits to the data. So, we first attempt to determine the amount of error inherent in our modeling, in order to assign more conservative uncertainties to the fitted parameters, as we will now describe. We first perform a steepest-descent $\chi^{2}$ minimization to obtain best-fitting parameters for the model. At this point, if the $\chi^{2}\sbr{red}$ value for the fit is greater than $1$, we assume that this is due to some error in the modeling, which we parametrize as $\sigma\sbr{m}$. We uniformly add this value in quadrature to the measured uncertainties in all radial bins, such that the adjusted $\chi^{2}\sbr{red} = 1$ for the best fit. We then repeat this process, finding new best-fit values and recalculating $\sigma\sbr{m}$ until convergence.

Since this procedure effectively increases the error in all radial bins, this process has the result of increasing the measured errors on all fitted parameters. If the model is initially a good fit ($\chi^{2}\sbr{red} \approx 1$) to the data, the increase is negligible, but if the model is a poor fit to the data, the estimated errors for the fitted parameters will be significantly increased. As such, this process allows us to place more conservative limits on our results, based on the quality of the model's fit to the data.

Additionally, since the model error is uniformly added to the errors in all radial bins, it prevents the fitting algorithm from over-weighting the fit to the high-radius bins, which otherwise have significantly lower errors, and thus typically contribute more to the $\chi^2$ value of the fit if the model isn't a perfect fit to the data.

For the models we tested, we typically found for the HDE samples that $\sigma\sbr{m} \lesssim 0.5 \msun/{\rm pc}^{2}$, which is $\lesssim 5\%$ of the measured lensing signal $\Delta\Sigma$. For the LDE samples, most fits were initially of $\chi^{2}\sbr{red} \approx 1$, and so no model error term was necessary. 

Once the model error is determined, we run an MCMC algorithm to help determine the errors of the fitted parameters. Since only the mass of satellite halos is relevant to us, we marginalize over all other parameters to get the mean value and errors for the satellite mass.

\section{Results and Analysis}

\label{sec:results}

\begin{figure*}
\centering
\includegraphics[scale=0.4]{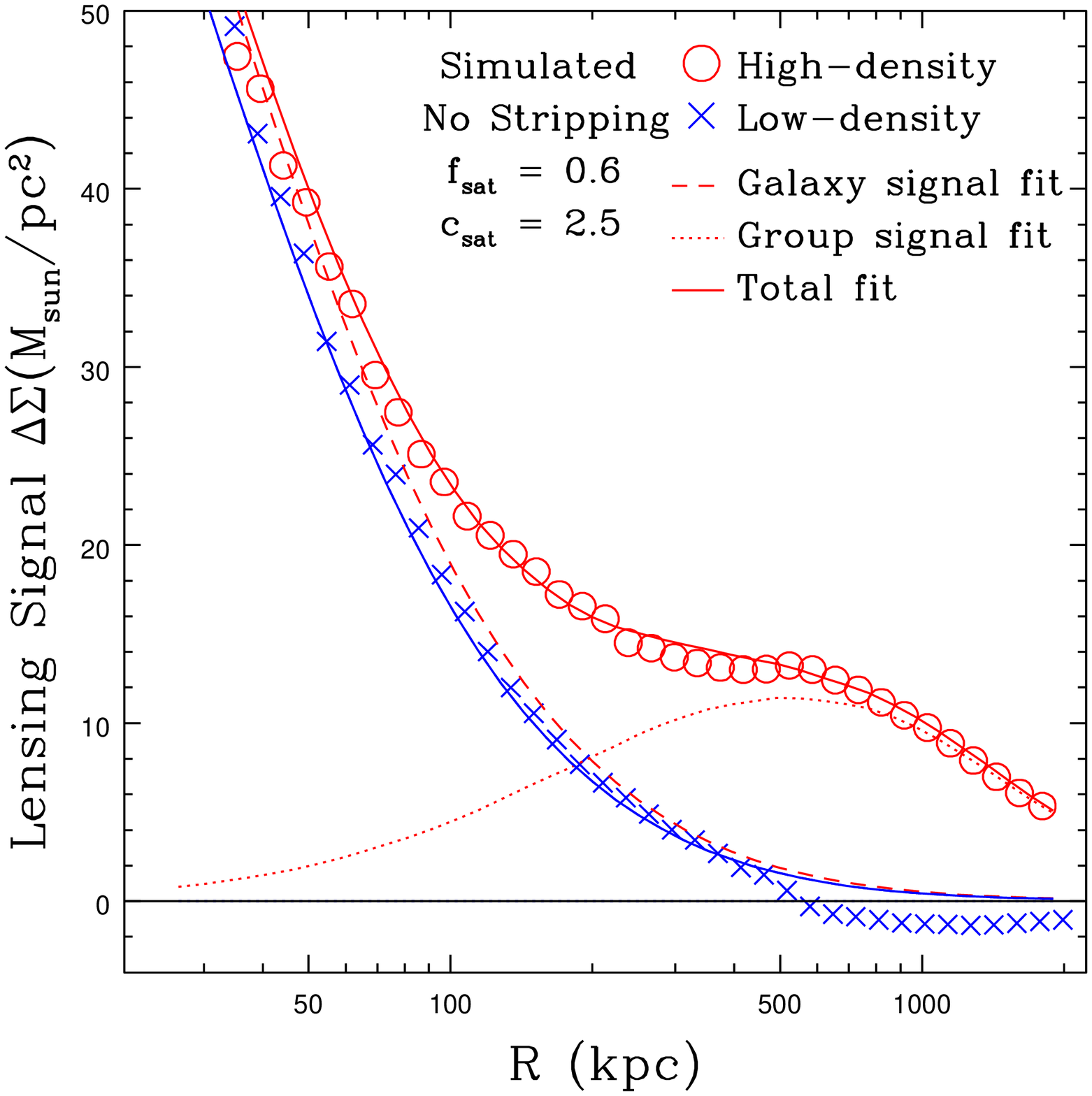}
\includegraphics[scale=0.4]{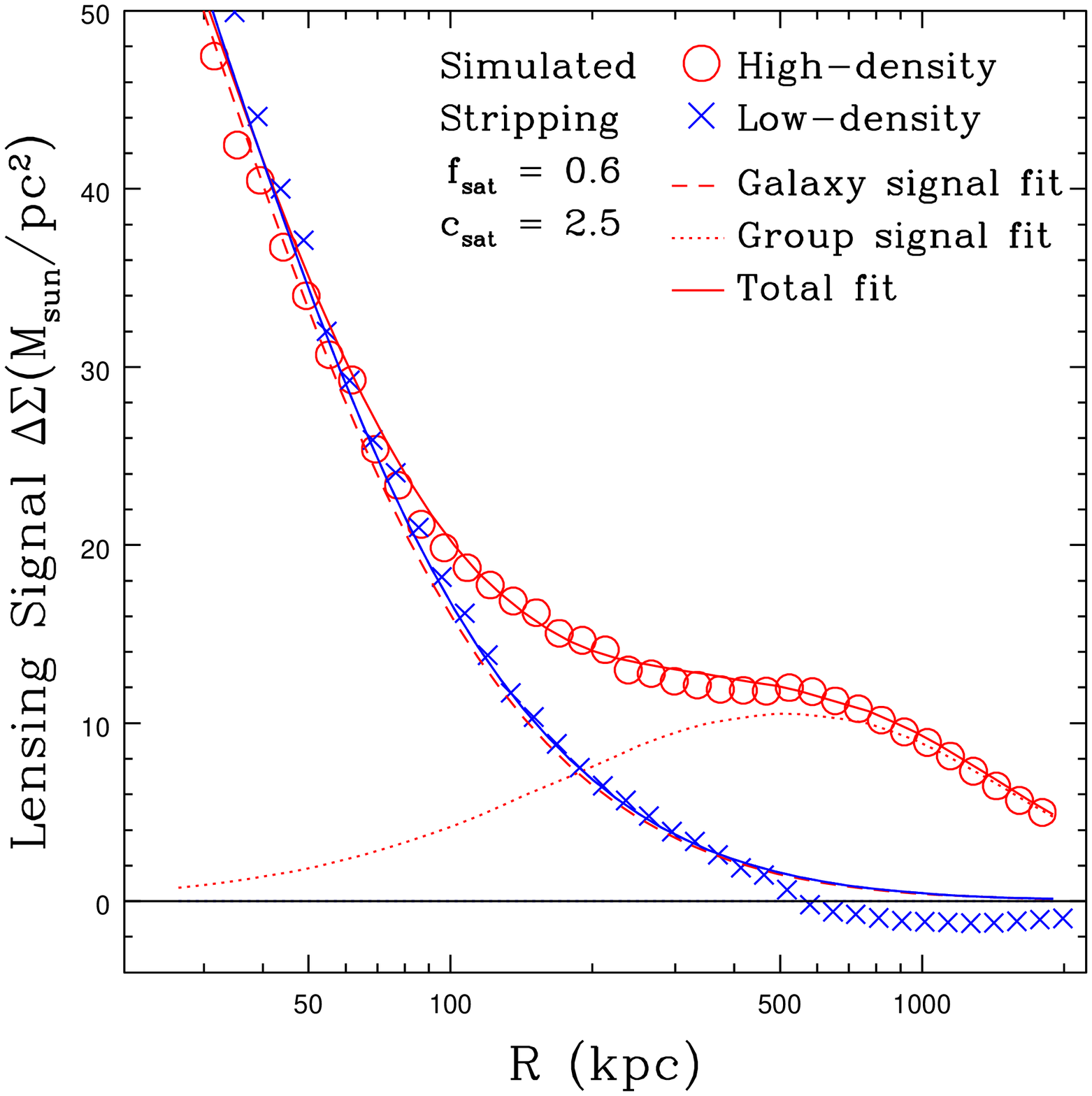}
\caption[]{Lensing signals and fits for simulated lensing data for the ``No Stripping'' (left) and ``Stripping'' (right) scenarios (see \secref{simdata}). The ``No Stripping'' scenario shows similar one-halo fits for the HDE and LDE samples, while the ``Stripping'' scenario shows a lower one-halo mass fit for the HDE sample than for the LDE sample. Error bars are not shown, as shape noise is not simulated for these datasets, and so the scatter is extremely small.}
\label{fig:millsig}  
\end{figure*}

\begin{table*}
\centering
\caption{Results of the fitting procedure when applied to simulated (top) and the CFHTLenS (bottom) lensing data in various stellar mass bins. All masses are in units of $10^{10}$~$\msun$. $\log{m}$ is the stellar mass bin. $f\sbr{sat}$ is the fraction of satellites we use for the fitting, based on data from the Millennium Simulation. $M\sbr{HDE}$ and $M\sbr{LDE}$ are the fitted one-halo masses for the HDE and LDE samples. $M\sbr{gr}$ is the fitted mass of the offset group halo term. R$_{M}$ is the ratio of $M\sbr{HDE}$ to $M\sbr{LDE}$. $\chi^{2}\sbr{red}$ is the reduced $\chi^2$ parameter without the model error term (see \secref{fitpro}) included (for 36 degrees of freedom; a value close to 1 is ideal).}
\begin{tabular}{| l l l l l l l l l l l l l l|}
 & & & \multicolumn{4}{|c|}{``No Stripping'' Model} & & \multicolumn{4}{c|}{``Stripping'' Model} \\
\hline
$\log{m}$ & $f\sbr{sat}$ & & $M\sbr{HDE}$ & $M\sbr{gr}$ & $M\sbr{LDE}$ & R$_{M}$ & & $M\sbr{HDE}$ & $M\sbr{gr}$ & $M\sbr{LDE}$ & R$_{M}$ \\ \hline
\phn\phd\phn9--9.5 & 0.53 & & \phn20 & 12000 & \phn21 & 0.95 & & \phn14 & 9800 & \phn19 & 0.74 \\
\phn9.5--10 & 0.60 & & \phn46 & 11000 & \phn41 & 1.12 & & \phn32 & 9700 & \phn39 & 0.83 \\
\phn\phd10--10.5 & 0.63 & & 140 & \phn7300 & 110 & 1.27 & & 110 & 7200 & 120 & 0.94 \\
10.5--11 & 0.48 & & 930 & \phn9600 & 650 & 1.43 & & 950 & 5900 & 660 & 1.44 \\

\hline
\end{tabular}
\begin{tabular}{| l l l l l l l l l|}
 & & & \multicolumn{5}{|c|}{CFHTLenS Data} \\
\hline
$\log{m}$ & f$\sbr{sat}$ & & $M\sbr{HDE}$ & $\chi^{2}\sbr{red,HDE}$ & $M\sbr{gr}$ & $M\sbr{LDE}$ & $\chi^{2}\sbr{red,LDE}$ & R$_{M}$ \\ [1ex]
\phn\phd\phn9--9.5 & 0.53 & & \phn\phn17.6 $\pm$ \phn\phn 4.8 & 2.31 & 20500 $\pm$ 2300 & \phn24.9 $\pm$ \phn\phn4.0 & 0.83 & 0.71$^{+0.25}_{-0.18}$ \\ [1ex]
\phn9.5--10 & 0.60 & & \phn\phn16.5 $\pm$ \phn\phn6.5 & 1.05 & 15060 $\pm$ \phn900 & \phn35.6 $\pm$ \phn\phn6.2 & 0.80 & 0.46$^{+0.25}_{-0.15}$ \\ [1ex]
\phd\phn10--10.5 & 0.63 & & \phn\phn67 \phd\phn $\pm$ \phn12 & 0.65 & 14550 $\pm$ \phn550 & \phn95 \phd\phn $\pm$ \phn11 & 0.90 & 0.70$^{+0.17}_{-0.12}$ \\ [1ex]
10.5--11 & 0.45 & & \phn287 \phd\phn $\pm$ \phn34 & 1.45 & 23100 $\pm$ 4000 & 239 \phd\phn $\pm$ \phn38 & 1.41 & 1.20$^{+0.30}_{-0.21}$ \\ [1ex]
\phd\phn11--11.5 & 0.45 & & 1090 \phd\phn $\pm$ 120 & 0.81 & 20300 $\pm$ 2000 & 530 \phd\phn $\pm$ 110 & 1.29 & 2.05$^{+0.65}_{-0.31}$ \\ [1ex]
\hline
\end{tabular}
\label{tab:fits}
\end{table*}

In this section, we present the results of the fits and discuss their implications. In \secref{simpred}, we discuss the predicted results from the simulations, for both the ``No Stripping'' and ``Stripping'' models. \secref{obsred} presents the main results of our analysis of the CFHTLenS dataset and discuss their implications. In \secref{altfits}, we discuss alternative interpretations of the data, and which of the one-halo mass, concentration, and truncation radius might plausibly contribute to the observed differences between the HDE and LDE samples. \secref{analysis} discusses potential systematic effects.

\subsection{Predictions from Simulations}
\label{sec:simpred}

\figref{millsig} shows plots of the best-fit models for the simulated catalogues from the Millennium Simulation, for both the ``Stripping'' and ``No Stripping models'' (described in \secref{simlens}), for galaxies with $10^{9} \msun < m < 10^{10.5} \msun$. The plot illustrates that in the ``No Stripping'' scenario, the measured lensing signals for the HDE and LDE samples are nearly identical at very small radii. Our algorithm does not work perfectly for this mass bin, and in the ``No Stripping'' scenario, it fits a one-halo mass to the HDE sample that is somewhat larger than the one-halo mass fitted to the LDE sample, while for the ``Stripping'' scenario, the fitted HDE one-halo mass is slightly lower than the fitted LDE one-halo mass.

Further comparisons of fitted one-halo masses for different mass bins can be seen in \tabref{fits} and \figref{fitscomp}. As can be seen there, for all mass bins $m < 10^{10.5} \msun$ with the ``Stripping'' model, as expected the fit yields a relatively lower one-halo mass for the HDE sample compared to the LDE sample than it does for the ``No Stripping'' model. Above $m = 10^{10.5} \msun$, however, the fitted masses in the ``Stripping'' and ``No Stripping'' scenarios are comparable. This is due to the fact that at high stellar masses, the fraction of galaxies in the HDE sample that are centrals increases rapidly (see \tabref{simstats}). Since mass stripped from satellites is added to the masses of central galaxies, then if too many central galaxies are included in the sample, stripping will have little or no net effect on the lensing signal.

The fitted group masses for the simulated data seen in \tabref{fits} are larger than the actual group masses by a factor of $\sim1.5$--$2$. Our tests have shown that this can occur when halos from a very broad range of masses are averaged together, as is the case here -- the lensing signal of an average of halos of varying mass is similar to the lensing signal of a single halo with a mass somewhat greater than the average of the sample. The fitted group masses for the CFHTLenS data are additionally observed to be a factor of $\sim2$ larger than the group masses for simulated data. This is not surprising, as the halo masses in the simulated data are extrapolated from the stellar masses of their constituent galaxies, and the distribution of stellar masses in the Millennium Simulation does not match the distribution in the CFHTLenS dataset.

These results from the simulations imply that with the CFHTLenS data, a comparison of the HDE and LDE fitted one-halo masses can be used as an indication of whether or not tidal stripping is occurring, but we must use a stellar mass upper limit of $\sim 10^{10.5} \msun$.

\subsection{Observational Results}
\label{sec:obsred}

\begin{figure}
\centering
\includegraphics[scale=0.4]{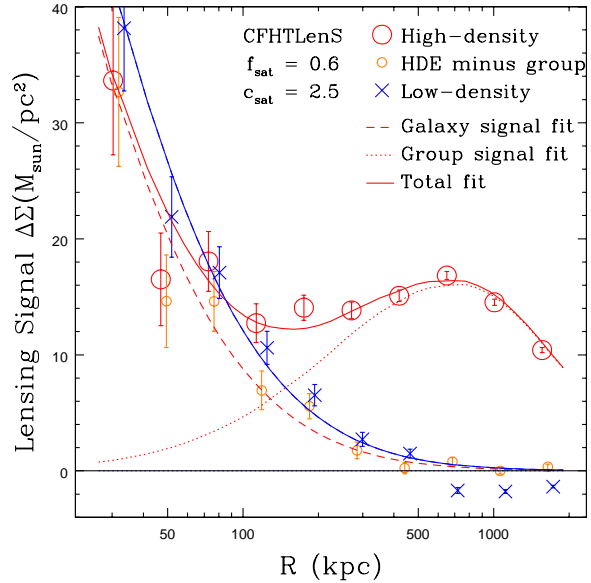}
\caption[]{Measured lensing signal and model fits for data from the CHFTLenS survey, including all galaxies with $10^{9} < m < 10^{10.5} \msun$. HDE (red) and LDE (blue) lensing signals and fits are illustrated, as well as the HDE data with the fitted offset-group-halo term subtracted off (orange). The dashed line shows the one-halo model fit to the HDE sample, and the dotted line shows the HDE offset-group-halo term. The one-halo mass fit for the HDE sample is found to be significantly lower than for the LDE sample.
}
\label{fig:widesig}  
\end{figure}

\figref{widesig} shows the lensing signals for the HDE and LDE samples taken from the CFHTLenS survey, including all galaxies with $10^{9} \msun < m < 10^{10.5}$~$\msun$, with the best-fit models plotted on top. For this broad mass bin, the fits show that the HDE one-halo term is lower than the LDE term, at $2.5\sigma$ significance ($p=0.0113$). However, this simple fit is not optimal. In part, this is because we are combining galaxies with greatly varying masses. The resultant lensing signal of this combination does not perfectly resemble the lensing signal of a single halo possessing the average mass of the sample, and the code compensates for this by fitting a higher $\sigma\sbr{m}$, which results in larger errors for the best fit.  

\begin{figure}
\includegraphics[scale=0.47]{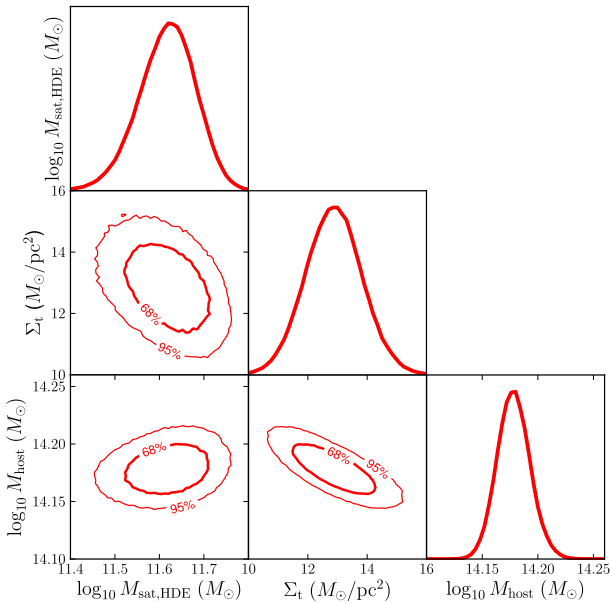}
\caption{Probability distribution functions and joint probability distribution functions for satellite mass M$\sbr{sat}$ (left column), host group mass M$\sbr{host}$ (bottom row), and surface density threshold $\Sigma\sbr{t}$ (middle row and middle column) for the fit of the lensing signal of all HDE galaxies with $10^9 \msun < m < 10^{10.5} \msun$.}
\label{fig:2dhists}
\end{figure}

\figref{2dhists} shows the likelihood distributions for the fitted satellite masses, host group mass, and surface density threshold for the HDE sample of galaxies with $10^9 \msun < m < 10^{10.5} \msun$. The plot shows that there is only a weak degeneracy of M$\sbr{sat}$ with the other two parameters, but there is a stronger degeneracy between M$\sbr{host}$ and $\Sigma\sbr{t}$. Nevertheless, when marginalized over the other parameters, M$\sbr{host}$ is very tightly constrained, and M$\sbr{sat}$ is reasonably constrained.

\begin{figure}
\centering
\includegraphics[scale=0.4]{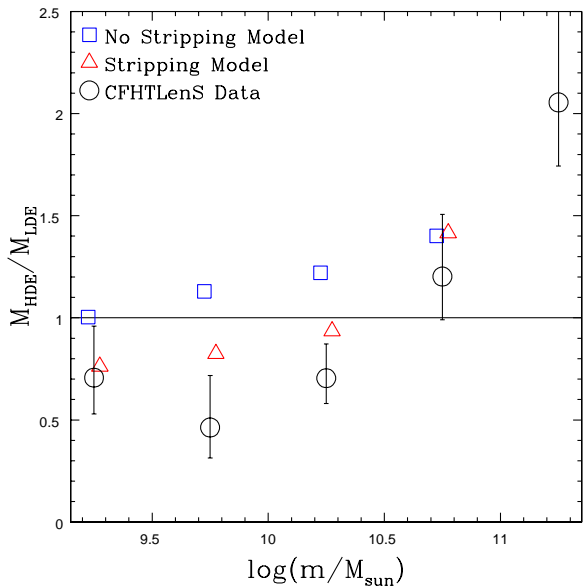}
\caption[]{A summary plot of the fitting results for various stellar mass bins. Shown are the ratios of the best-fit one-halo mass for the HDE and LDE samples, for both the Stripping and No Stripping simulations, and for the CFHTLenS data.}
\label{fig:fitscomp}  
\end{figure}

We can more carefully analyze the data by splitting the galaxy sample into smaller stellar mass bins. \figref{fitscomp} shows the results of this analysis for both simulated and the CFHTLenS data, with the ratio of the fitted one-halo mass for the HDE sample to that of the LDE sample plotted against the galaxies' stellar masses. Simulated data is not available for all mass bins plotted due to limitations of the Millenium catalogue. The simulated data demonstrates that for sufficiently high stellar mass bins, the ``Stripping'' HDE mass becomes comparable to or greater than the ``No Stripping'' HDE mass. This is due to the fact that, at high stellar masses, the P3-identified HDE sample contains a large number of centrals. When tidal stripping is present, mass is transferred from satellites to centrals, increasing their mass. When centrals make up a large enough fraction of the sample, the sample shows an increase in mass under the effects of tidal stripping. Therefore, we restrict the analysis to bins with $m < 10^{10.5} \msun$,  where the ``Stripping'' scenario predicts lower fitted mass than the ``No Stripping'' scenario.

Details of the fits to CFHTLenS data for different stellar mass bins are shown in \tabref{fits}. The goodness-of-fit is comparable to previous galaxy-galaxy lensing studies. Specifically, the $\chi^{2}\sbr{red}$ values for our fits (which are calculated before the inclusion of the model error term, see \secref{fitpro}) are similar to the full halo model fits of \citet{VelUitHoe12}: their $\chi^{2}\sbr{red}$ values varied from 0.5--2 for different stellar mass bins, whereas ours vary from 0.6--2.3.

If only the three stellar mass bins with $10^{9} < m < 10^{10.5} \msun$ are used, we obtained a weighted mean ratio of HDE one-halo mass to LDE one-halo mass of $0.65\pm0.12$. If we assume that this ratio is indicative of the retained mass after stripping, and assume the sample contains $\sim60\%$ satellites, then we can extrapolate that for a sample of $100\%$ satellites, the mean retained mass fraction will be $\sim0.41\pm0.19$, which is consistent with the mean retained mass fraction of $0.40$ we measured from the simulated data.

Note that at face value, our result suggests less mass reduction in HDE environments than the factors of 2--5 found for the $\sim L_*$ satellites of the rich cluster Cl 0024+16 found by \cite{NatKneSma09}.  There are several key differences between these samples, however; in particular, our satellites have lower stellar mass and our satellites inhabit lower mass host haloes than the rich cluster studied by \cite{NatKneSma09}. 

These combined results reject the results of the simulated ``No Stripping'' model at $4.1 \sigma$ ($p<0.0001$), reject $M\sbr{HDE} = M\sbr{LDE}$ at $2.9 \sigma$ ($p=0.0039$), and are consistent with the simulated ``Stripping'' model at $1.8 \sigma$ ($p=0.0651$). This near-rejection of the ``Stripping'' model may indicate that this model underestimates the amount of tidal stripping which occurs in reality, or it might indicate that some effect other than tidal stripping (such as a difference in star formation histories dependant on environment) is contributing to the observed signal. Additionally, while our results have high statistical significance, they do not rule out the possibility of systematic errors resulting in a spurious detection. We investigate the possibility of such a spurious detection in \secref{analysis}.

\subsection{Alternative Fits}
\label{sec:altfits}

In the previous subsection, we made a number of assumptions relating one-halo mass, concentration, and tidal radius (see \eqref{cfm}). This allowed us to compare only the one-halo mass between the HDE and LDE samples, to determine if there was a difference in their lensing signals. However, this doesn't tell us what variations in mass, concentration, and tidal radius might be causing this difference. In this section we investigate alternate means of fitting the lensing signals to determine which of these parameters might differ between the HDE and LDE samples.

\subsubsection{Truncation Radius}
\label{sec:trunc}

An alternative  fit with one free one-halo parameter is to fit the HDE model with the same one-halo mass as the LDE sample, but with a lower tidal (truncation) radius for its halo, as is predicted by typical models of tidal stripping. For the single broad stellar mass bin $10^9 < m < 10^{10.5} \msun$, we fit a tidal radius of $r\sbr{tidal}/r\sbr{200} = 0.26 \pm 0.14$ for the HDE galaxies. This corresponds to a typical tidal radius of $40 \pm 21$~kpc and a retained mass fraction of $0.43 \pm 0.18$ for a satellite galaxy with $M = 5.9 \times 10^{11} \msun$ and $c=8.5$. However, this model is a marginally poorer fit compared to the default method of fitting one-halo mass: $\chi^{2}\sbr{red} = 1.08$ for fitting tidal radius, compared to $\chi^{2}\sbr{red} = 0.99$ for fitting mass. Our data is thus consistent with both interpretations, and we are unable to discern between them.

\subsubsection{One-halo Mass and Concentration}
\label{sec:1hc}

\begin{figure}
\centering
\includegraphics[scale=0.5]{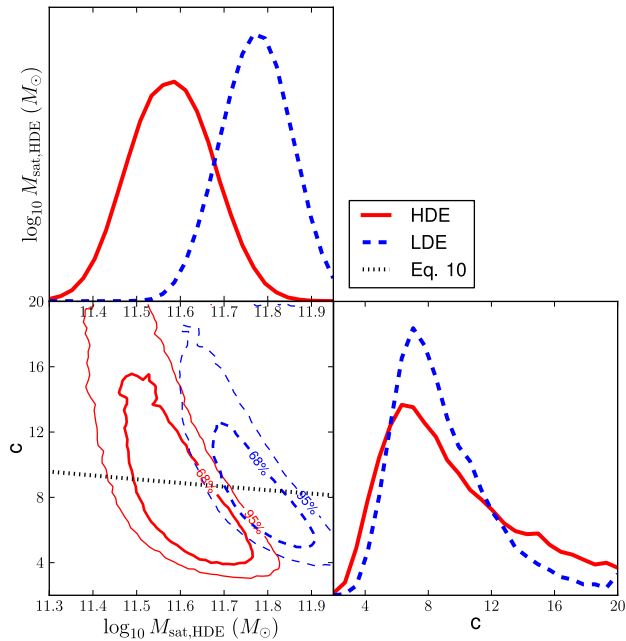}
\caption[]{Lower left: Joint probability distribution function for one-halo mass and concentration for the HDE (red solid lines) and LDE (blue dashed lines) galaxy samples. The dotted line shows the relation between mass and concentration given by \eqref{cfm}, which was used for previous analysis. Upper left: Probability distribution function for one-halo mass, marginalized over concentration. Lower right: Probability distribution function for concentration, marginalized over one-halo mass.}
\label{fig:contourmc}  
\end{figure}

It is also possible that the observed difference in lensing signals between HDE and LDE galaxies could be due to a difference in concentration, rather than a difference in mass. Halos with lower concentrations will have lower lensing signals at small radii, and somewhat higher lensing signals at large radii.
To test whether the observed results could be due to a change in concentration, we reran the analysis, leaving concentration as a free parameter. The resultant joint probability distribution function for this analysis is shown in \figref{contourmc}. As this plot shows, while there is a difference in the mass PDFs for the HDE and LDE samples, there is no evidence for a difference in concentration between the two samples. We therefore favour the interpretation that the measured difference in lensing signals between the HDE and LDE samples is due to the HDE halos being less massive than LDE halos, but we are unable to rule out the possibility that the two samples have the same mass, but the HDE sample is less concentrated.

\subsection{Analysis}
\label{sec:analysis}

Although the results are statistically significant, it is nevertheless possible that systematic errors have entered our analysis. In this section, we discuss various possible systematic errors that may affect the results.

\subsubsection{Stellar Mass Biases}
\label{sec:bias}

In principle, if there is a relative bias in the estimates of stellar mass between red and blue galaxies, and if the HDE and LDE samples contain different fractions of red and blue galaxies, we could get a spuriously positive detection of tidal stripping with our method. It is difficult to completely rule out the possibility of a relative bias in stellar mass, but it can be investigated by comparing the distributions of stellar mass to other published distributions. Unlike the CFHTLenS data, the WIRDS \citep{BieHudMcC12} data are based on optical photometry supplemented by deep infrared images. We would therefore expect the stellar masses in WIRDS to be more accurate. A comparison of stellar mass estimates between CFHTLenS and WIRDS \citep{VelUitHoe12} shows that, if we assume WIRDS stellar masses to be more accurate, then CFHTLenS stellar masses may indeed be slightly biased, with CFHTLenS red galaxy stellar masses $\sim0.05$--$0.1$ dex too low and the opposite for CFHTLenS blue galaxies. Since we are selecting by CFHTLenS stellar mass, and because red galaxies are more common in high-density environments, this implies that the true mean stellar mass in HDE regions is actually slightly larger than that in the ``matched'' LDE regions. Correspondingly, in absence of stripping, we would expect the recovered subhalo masses to be larger.  Since we find them to be smaller, this effect is in the wrong sense to explain our stripping detection. 

In any case, the effect is small. As can be seen in \tabref{obsstats}, the maximum difference in the fraction of red galaxies between the HDE and LDE samples is in fact only $10\%$, in the $10^{9.5}$~$\msun < m < 10^{10}$~$\msun$ and $10^{10}$~$\msun < m < 10^{10.5}$~$\msun$ bins. This, combined with the above estimates of the relative bias, allows us to put an upper limit on this effect of $\sim5\%$.

\subsubsection{Modeling Inaccuracies}
\label{sec:modelinac}

We have attempted to account for inaccuracies in the models through the inclusion of a ``model error'' term, but this does not account for all possible errors in modeling that might arise. Notably, for the HDE sample, there is a weak degeneracy between the fitted one-halo term and the other fitted parameters, as well as with the fixed parameters: the fraction of galaxies which are satellites ($f\sbr{sat}$) and the concentration of satellites ($c\sbr{sat}$). We marginalize over the other fitted parameters to estimate the mean one-halo mass and its error, but errors in $f\sbr{sat}$ and $c\sbr{sat}$ would persist through the analysis as systematic errors.

\begin{table}
\centering
\caption{Results of varying the fixed parameters ($f\sbr{sat}$ and $c\sbr{sat}$) in the fitting procedure, using a sample of galaxies with $10^{9}$~$\msun < m < 10^{10.5}$~$\msun$. All masses are in units of $10^{10}$~$\msun$.}
\begin{tabular}{| l l l l l l l l l l l l l |}
$f\sbr{sat}$ & $c\sbr{sat}$ & & $M\sbr{HDE}$ & R$_{M}$ & $M\sbr{group}$ & $\chi^{2}\sbr{red}$ \\ \hline
0.4 & 2.5 & & 44  $\pm$ 6  & 0.76 & 23000 & \phn1.17 \\ 
\textbf{0.6} & \textbf{2.5} & & \textbf{38}  $\mathbf{\pm}$ \textbf{5}  & \textbf{0.65} & \textbf{15000} & \textbf{\phn0.99} \\
0.8 & 2.5 & & 33  $\pm$ 5  & 0.56 & 12000 & \phn1.22 \\ 
0.4 & 5.0 & & 35  $\pm$ 4  & 0.59 & 25000 & \phn1.08 \\ 
0.6 & 5.0 & & 25  $\pm$ 4  & 0.43 & 17000 & \phn1.93 \\ 
0.8 & 5.0 & & 20  $\pm$ 4  & 0.33 & \phn9100 & 12.09 \\ 
\hline
\end{tabular}
\label{tab:modelanalysis}
\end{table}

\begin{figure*}
\centering
\includegraphics[scale=0.4]{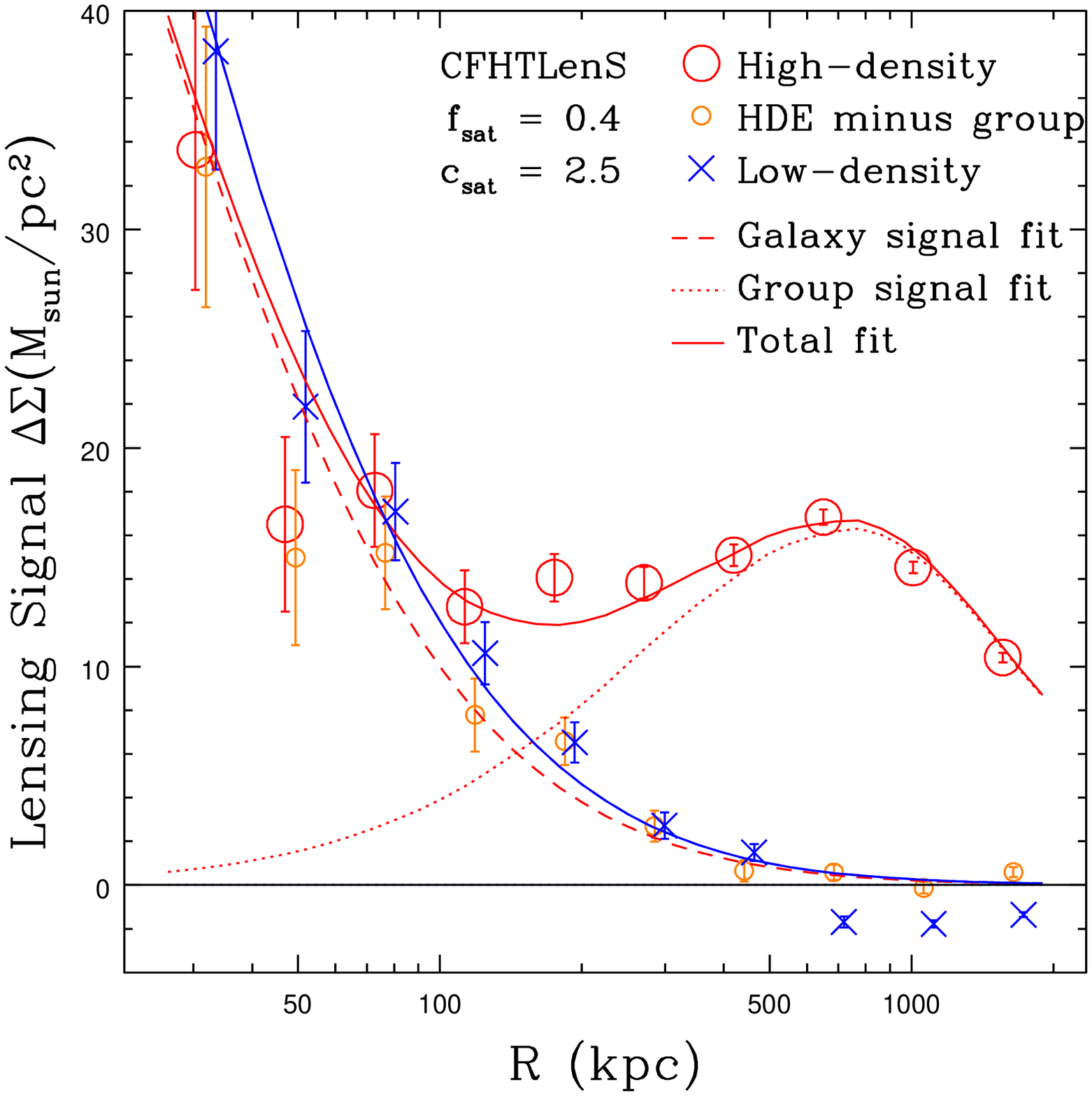}
\includegraphics[scale=0.4]{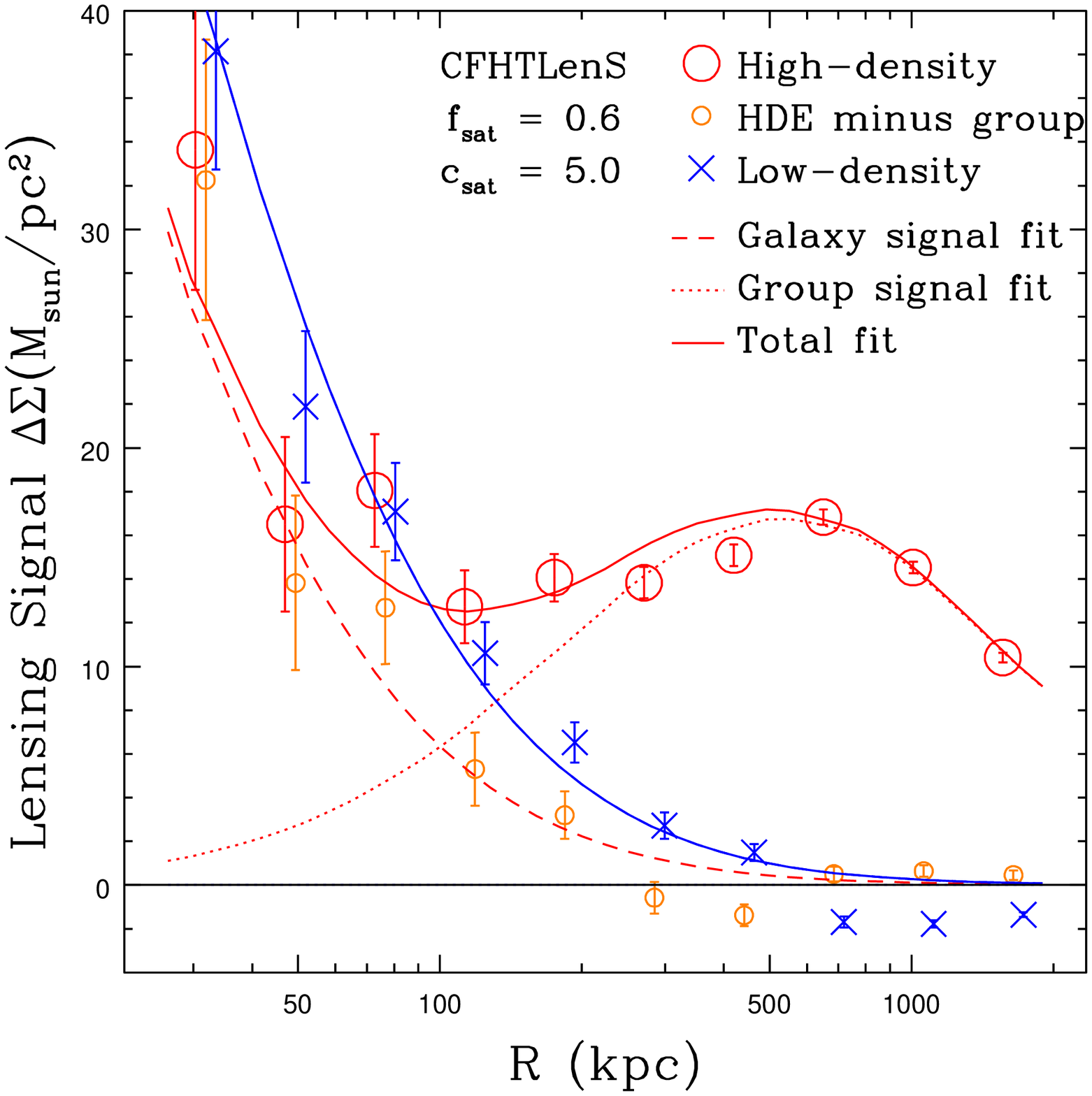}
\caption[]{As \figref{widesig}, except with the fraction of satellites fixed to 0.4 (left), and the concentration of satellites fixed to 5.0 (right) for the fitting of the HDE signal.
}
\label{fig:varscf}  
\end{figure*}

To assess the potential impact of errors in the fraction of satellites and satellite concentration, we reran the analysis varying these parameters, to observe how much the results changed. The results plotted here use a sample of all galaxies with $m<10^{10.5} \msun$, weighted by estimated halo mass. The results of this analysis are presented in \tabref{modelanalysis}, and sample plots of the fits can be observed in \figref{varscf}. We allow the fraction of satellites to vary from $0.4$ to $0.8$, and we test increasing the concentration of satellites to 5.0, which is comparable to the concentration of the dark matter halos of galaxy groups.

As compared with the default case ($f\sbr{sat}=0.6$ and $c\sbr{sat}=2.5$), the only variation of the parameters that results in increasing the fitted HDE mass is when the fraction of satellites is decreased. If the tested fraction of 0.4 were the case in reality, the significance of the detection would decrease to $\sim2.1 \sigma$. However, this particular satellite fraction results in a marginally poorer fit, as evidenced by the increased $\chi^{2}\sbr{red}$ value.

Alternatively, if the actual concentration of satellites within groups is higher, the fitted satellite mass will be lower. A fit with $c\sbr{sat} = 5.0$ yields a marginally poorer $\chi^{2}\sbr{red}$ value, but we cannot rule out this scenario. If this is the case in reality, it would strengthen the significance of the detection to $\sim4.4 \sigma$.

\begin{table}
\caption{Fitted values of $\Sigma\sbr{t}$ for various stellar mass bins, in units of $\msun/{\rm pc}^2$.}
\begin{tabular}{| l l l l l l l l l l l l l |}
$\log{m}$ & $\Sigma\sbr{t}$ \\ \hline
\phn\phd\phn9--9.5 & 42 $\pm$ 2 \\
\phn9.5--10 & 34 $\pm$ 3 \\
\phd\phn10--10.5 & 44 $\pm$ 3 \\
10.5--11 & 39 $\pm$ 6 \\
\phd\phn11--11.5 & 36 $\pm$ 7 \\
\hline
\end{tabular}
\label{tab:sigmatest}
\end{table}

It is also important to look at the impact of the density threshold term, $\Sigma\sbr{t}$. Although this term was found to be necessary in the simulated data to provide a reasonable fit when the fraction of satellites is known exactly, it is not certain that this is an accurate description. We can investigate this matter by looking at how the fitted value of $\Sigma\sbr{t}$ varies in the fits, as can be seen in \tabref{sigmatest}. If we were modeling everything perfectly, we would expect to see $\Sigma\sbr{t}$ being roughly constant for all stellar mass bins, as the P3 algorithm is blind to galaxy mass, and so there is no reason the threshold $\Sigma\sbr{t}$ for S/N $>$ 2 regions should vary with the mass of satellites. In fact, we observe $\Sigma\sbr{t}$ to remain roughly constant at $\sim 40 \msun/{\rm pc}^{2}$, which is consistent with the hypothesis that we are modeling it reasonably.

Other modeling inaccuracies may also affect the results. The model for the HDE sample neglects the contribution of the two-halo term (the contribution of nearby groups and field galaxies) to the lensing signal around satellites. We tested the implications of this with a rough model of the two-halo term, and it resulted in decreasing the fitted HDE satellite mass. Therefore, any possible systematic error in our results from this effect would be in the wrong sense to contribute to a spurious detection.

Additionally, the model for the LDE sample neglects the contribution of the local underdensity (see \secref{under}) to the lensing signal. Proper handling of this term would likely result in a slight increase in the fitted LDE mass, which would increase the significance of the detection.

\section{Conclusions}

\label{sec:conc}

Previous lensing analyses of the environmental dependence of satellite halo masses \citep{ManSelKau06,HoeVel11} have revealed the difficulty of detecting such an effect in samples where the satellites are predominantly expected to lie in groups. The analysis here improves on previous work by using the much deeper data provided by the CFHTLenS sample to better constrain the lensing signal around low-mass satellite galaxies in high-density environments.

Using photometric redshifts we divide galaxies in high-density and low-density environment subsamples that are matched in stellar mass. We have found a significant difference in their halo masses. Our analysis shows a highly significant ($4.4\sigma$) rejection of the simulated ``No Stripping'' model, and a significant ($3.2\sigma$)  rejection of the simple ``null hypothesis'' that there is no difference in the halo properties of HDE and LDE galaxies, for galaxies in a broad range of group masses. This difference is most likely due to tidal stripping of dark matter, and if so, this analysis represents the first detection of tidal stripping in a selection of galaxies that do not all reside within galaxy clusters.

We argue that these results are unlikely to be due to systematic errors in our methodology, as most suspected systematic errors would tend to bias one against a detection of stripping.

The mean ratio of fitted mass for the high-density environment sample to that of the low-density environment sample is $\sim0.65\pm0.12$. Since the HDE galaxy sample consists of only $\sim 60$\% satellites, the retained mass fraction for a pure satellite sample would be considerably lower: $\sim 0.41\pm0.19$. We can alternatively model this as the HDE satellites being tidally stripped at a typical radius of $r\sbr{tidal}/r\sbr{200} = 0.26 \pm 0.14$. This corresponds to a typical tidal radius of $40 \pm 21$~kpc and a retained mass fraction of $0.43 \pm 0.18$ for a satellite galaxy with $M = 5.9 \times 10^{11} \msun$ and $c=8.5$.

Further work will be necessary to confirm these results, and to analyze the dependence of stripping on both group mass and on the satellite's location within the group. 

\section*{Acknowledgments}

Based on observations obtained with MegaPrime/MegaCam, a joint project of CFHT and CEA/DAPNIA, at the Canada-France-Hawaii Telescope (CFHT) which is operated by the National Research Council (NRC) of Canada, the Institut National des Sciences de l'Univers of the Centre National de la Recherche Scientifique (CNRS) of France, and the University of Hawaii. This work is based in part on data products produced at TERAPIX and the Canadian Astronomy Data Centre as part of the Canada-France-Hawaii Telescope Legacy Survey, a collaborative project of NRC and CNRS.

This work was made possible by the facilities of the Shared Hierarchical Academic Research Computing Network (SHARCNET:www.sharcnet.ca) and Compute/Calcul Canada.

BRG acknowledges useful conversations with James Taylor.
MJH acknowledges support from the Natural Sciences and Engineering Research Council of Canada (NSERC).
TE is supported by the Deutsche Forschungsgemeinschaft through project ER 327/3-1 and the Transregional Collaborative Research Centre TR 33 - "The Dark Universe".
HH is supported by the Marie Curie IOF 252760, a CITA National Fellowship,
and the DFG grant Hi 1495/2-1.
LVW acknowledges support from the Natural Sciences and Engineering Research Council of Canada (NSERC) and the Canadian Institute for Advanced Research (CIfAR, Cosmology and Gravity program).
CH acknowledges support from the European Research Council under the EC FP7 grant number 240185.
H. Hoekstra acknowledges support from  Marie Curie IRG grant 230924, the Netherlands Organisation for Scientific Research (NWO) grant number 639.042.814 and from the European Research Council under the EC FP7 grant number 279396.
TDK acknowledges support from a Royal Society University Research Fellowship.
YM acknowledges support from CNRS/INSU (Institut National des Sciences de l'Univers) and the Programme National Galaxies et Cosmologie (PNCG).
CB is supported by the Spanish Science MinistryAYA2009-13936 Consolider-Ingenio CSD2007-00060, project2009SGR1398 from Generalitat de Catalunya and by the European Commission’s Marie Curie Initial Training Network CosmoComp (PITN-GA-2009-238356).
LF acknowledges support from NSFC grants 11103012 \& 10878003, Innovation Program 12ZZ134 and Chen Guang project 10CG46 of SMEC, and STCSM grant 11290706600.
MK is supported in parts by the DFG cluster of excellence `Origin and Structure of the Universe'. 
BR acknowledges support from the European Research Council in the form of a Starting Grant with number 24067. 
TS acknowledges support from NSF through grant AST-0444059-001, SAO through grant GO0-11147A, and NWO.
ES acknowledges support from the Netherlands Organisation for Scientific Research (NWO) grant number 639.042.814 and support from the European Research Council under the EC FP7 grant number 279396. 
MV acknowledges support from the Netherlands Organization for Scientific Research (NWO) and from the Beecroft Institute for Particle Astrophysics and Cosmology.

{\small Author Contributions: All authors contributed to the
  development and writing of this paper.  The authorship list reflects
  the lead authors of this paper (BG, MJH) followed by two
  alphabetical groups.  The first alphabetical group includes key
  contributers to the science analysis and interpretation in this
  paper, the founding core team and those whose long-term significant
  effort produced the final CFHTLenS data product.  The second group
  covers members of the CFHTLenS team and other collaborators who made a contribution to the
  project and/or this paper. CH and LVW co-led the CFHTLenS
  collaboration.}

\bibliographystyle{mn2e}
\bibliography{mjh}

\bsp

\label{lastpage}

\end{document}